\documentclass[reprint,NumberedRefs]{JASA_mod}

\usepackage{mathrsfs}

\newtheorem{cor}{Corollary}
\newtheorem{prp}{Proposition}

\begin{document}

\title[Reproducing Kernel Based Beamformer]{Beamforming in the Reproducing Kernel Domain Based on Spatial Differentiation}


\author{Takahiro Iwami}
\email{iwami@design.kyushu-u.ac.jp}
\affiliation{Graduate School of Design, Kyushu University, 4-9-1 Shiobaru, Minamiku, Fukuoka, 815-8540, Japan}

\author{Naohisa Inoue}
\affiliation{Faculty of Design, Kyushu University, 4-9-1 Shiobaru, Minamiku, Fukuoka, 815-8540, Japan}

\author{Akira Omoto}
\affiliation{Faculty of Design, Kyushu University, 4-9-1 Shiobaru, Minamiku, Fukuoka, 815-8540, Japan}




\begin{abstract}
This paper proposes a novel beamforming framework in the reproducing kernel domain, derived from a unified interpretation of directional response as spatial differentiation of the sound field.
By representing directional response using polynomial differential operators, the proposed method enables the formulation of arbitrary beam patterns including non-axisymmetric.
The derivation of the reproducing kernel associated with the interior fields is mathematically supported by Hobson's theorem, which allows concise analytical expressions.
Furthermore, the proposed framework generalizes conventional spherical harmonic domain beamformers by reinterpreting them as spatial differential operators, thereby clarifying their theoretical structure and extensibility.
Three numerical simulations conducted in two-dimensional space confirm the validity of the method.
\end{abstract}


\maketitle


\section{\label{sec:1} INTRODUCTION}
Beamforming techniques have been widely applied in various fields such as acoustic signal processing, communications, and medical ultrasound imaging. Owing to the miniaturization of microphone elements and advances in multichannel recording technology, beamforming has become increasingly practical in recent years.
Historically, the development of beamforming began with line arrays using the delay-and-sum method and the minimum variance distortionless response (MVDR) method \cite{van2002,capon1969}.
More recently, axisymmetric beamformers formulated in the sp
herical harmonic domain \cite{meyer2002,boaz2015} have been widely used, enabling more flexible microphone configurations and beam patterns.
Spherical harmonic domain beamformers have been shown to be highly effective in practical applications, owing to the suitability of spherical harmonics as basis functions for representing sound fields.
Meanwhile, several studies have proposed the use of reproducing kernels (RKs) as effective, though not strictly orthonormal, basis functions for sound field representation \cite{ueno2018,iwami2023jasa,iwami2024jasa}.

Beamforming can also be interpreted as a technique that synthesizes a desired receiver directivity from multiple sensing elements.
Although receiver directivity is often treated as an angular function \cite{bauer1987,williams1999}, several studies \cite{balmages2007,poletti2005} have formulated simple directional responses, such as cardioid patterns, as spatial derivatives with respect to the radial coordinate.
Furthermore, axisymmetric directional responses have been achieved using higher-order radial differential operators \cite{kashiwazaki2022}.
These studies collectively suggest a close relationship between spatial differentiation and directional response.
In Kashiwazaki \textit{et al.} \cite{kashiwazaki2022}, axisymmetric directional responses were modeled by applying spatial differentiation along the radial direction, thereby realizing higher-order directional sensitivity. 
In contrast, Bilbao \textit{et al.} \cite{bilbao2019i,bilbao2019l} demonstrated that both directional radiation and reception can be represented as inner products in the spherical-harmonic domain by applying spatial differential operators determined by the spherical harmonics to the spherical-harmonic expansion of the sound field. 
These approaches collectively reveal a close connection between spatial differentiation and directivities.

In this study, we show that general directional reception can be theoretically regarded as spatial differentiation of the sound field using polynomial differential operators, and based on this perspective, we propose a beamformer formulation in the RK domain.
The proposed beamformer allows arbitrary beam formation including non-axisymmetric for any microphone configuration and spatial dimension, providing greater flexibility compared with conventional methods.
Moreover, the proposed framework enables the explicit specification of both directivity and reception position, which is further exploited in an application example of directional sound-field extraction.
In addition, we reinterpret axisymmetric beamformers in the spherical harmonic domain as spatial differentiation operators and demonstrate their extensibility to non-axisymmetric cases.

The validity of the proposed method is verified through three numerical experiments conducted in two-dimensional space with randomly distributed microphones and directional responses.

In this paper, the time dependence is assumed as $\exp(-\mathrm{i}\omega t)$.

\subsection{\label{subsec:1.2} Notation and Preliminaries}
\begin{itemize}

\item A multi-index and its operations are defined as follows \cite{grochenig2001}.
A multi-index is denoted by $\alpha = (\alpha_{1}, \ldots, \alpha_{d}) \in \mathbb{Z}_{+}^{d}$, and the following operations are associated with it:
\begin{align}
    |\alpha| &:= \sum_{i=1}^{d} \alpha_{i}, \quad \alpha! := \prod_{i=1}^{d} \alpha_{i}!, \notag \\
    \bm{r}^{\alpha} &:= \prod_{i=1}^{d} x_{i}^{\alpha_{i}}, \quad D^{\alpha} := \frac{\partial^{|\alpha|}}{\partial x_{1}^{\alpha_{1}} \cdots \partial x_{d}^{\alpha_{d}}}. \notag
\end{align}

\item Let $\mathscr{P}_{\nu}$ denote the space of homogeneous polynomials of degree $\nu$ on $\mathbb{R}^{d}$, and let $\mathscr{H}_{\nu}$ denote the space of homogeneous harmonic polynomials of degree $\nu$ on $\mathbb{R}^{d}$.

\item Let $\mathscr{Y}_{\nu}$ denote the space consisting of functions obtained by restricting the domain of homogeneous harmonic polynomials of degree $\nu$ on $\mathbb{R}^{d}$ to the unit sphere $S^{d-1}$.
In this paper, a complete orthonormal basis of $\mathscr{Y}{\nu}$, denoted by ${ Y_{\nu}^{\mu} }$, is referred to as the spherical harmonics.

\item For any polynomial $f(\bm{r}) = \sum_{\alpha \in \mathbb{Z}_{+}^{d}} c_{\alpha} \bm{r}^{\alpha}$, the corresponding constant-coefficient differential operator $f(D)$ is defined as
\begin{equation}
f(D) = \sum_{\alpha \in \mathbb{Z}_{+}^{d}} c_{\alpha} D^{\alpha}.
\end{equation}
When considering partial derivatives with respect to functions of two or more variables, the derivative with respect to the $i$-th variable is denoted by $\partial_{i}$ instead of $D$.
If it is necessary to explicitly indicate the variable being differentiated, the notation $\partial_{\bm{r}}$ is used.
\end{itemize}

\section{\label{sec:2} Interpretation of Directional Reception and Sound Field Reconstruction}
\subsection{\label{subsec:2.1} Directional Reception Model Based on Polynomial Differential Operators}
Consider the sound field in the frequency domain, where $c$ denotes the speed of sound, $\omega$ the angular frequency, and $k
:=\omega/c$ the wavenumber.
The classical directional response of a sensor is defined as its sensitivity $\zeta:S^{d-1} \times \mathbb{R}_{+} \rightarrow \mathbb{C}$ to plane wave incident from direction $\bm{\vartheta} \in S^{d-1}$.
That is, a sensing element located at position $\bm{r}_{\mathrm{r}}$ outputs a signal obtained by transforming an incident plane wave of angular frequency $\omega$ arriving from direction $\bm{\vartheta} \in S^{d-1}$:
\begin{equation} \label{eq:conventional_receiver_directivity}
    \mathrm{e}^{- \mathrm{i} k \bm{\vartheta} \cdot \bm{r}_{\mathrm{r}}} \mapsto \zeta(\bm{\vartheta}, \omega) \mathrm{e}^{- \mathrm{i} k \bm{\vartheta} \cdot \bm{r}_{\mathrm{r}}}.
\end{equation}
Here, $\zeta$ is referred to as the receiver directivity function of the sensing element.
For simplicity, the dependence on the angular frequency $\omega$ will be omitted.

In general, an interior sound field can be represented by the plane-wave expansion form \cite{iwami2024jasa}:
\begin{equation} \label{eq:plane_wave_expansion}
    \hat{p}(\bm{r}) = \int_{S^{d-1}} \tilde{P}_{\mathrm{b}}\left( k \bm{\vartheta} \right) \mathrm{e}^{-\mathrm{i} k \bm{\vartheta} \cdot \bm{r}} \mathrm{d} \bm{\vartheta}.
\end{equation}
When this sound field is received by a sensor with a directivity function $\zeta$ located at $\bm{r}_{\mathrm{r}}$, its output signal is given, from Eq.~(\ref{eq:conventional_receiver_directivity}), as
\begin{equation} \label{eq:directional_input}
    \hat{s}(\bm{r}_{\mathrm{r}}) = \int_{S^{d-1}} \tilde{P}_{\mathrm{b}}\left( k \bm{\vartheta} \right) \zeta(\bm{\vartheta}) \mathrm{e}^{-\mathrm{i} k \bm{\vartheta} \cdot \bm{r}_{\mathrm{r}}} \mathrm{d} \bm{\vartheta}.
\end{equation}
Since this expression involves no far-field assumption, it can be regarded as a general formulation of directional reception for arbitrary sound fields.

The space of square-integrable functions on the unit sphere, $L^{2}(S^{d-1})$ can be expressed as the algebraic direct sum of the homogeneous polynomials restricted to $S^{d-1}$, that is,
\begin{equation}
    L^{2}(S^{d-1}) = \sum_{\nu=0}^{\infty} {}^{\oplus} \mathscr{P}_{\nu} |_{S^{d-1}}.
\end{equation}
Hence, assuming that $\zeta \in L^{2}(S^{d-1})$, the directivity function can be written using a multi-index $\alpha$ as
\begin{equation} \label{eq:directivity_pol}
    \zeta(\bm{\vartheta}) = \sum_{\alpha \in \mathbb{Z}_{+}^{d}} c_{\alpha} \bm{\vartheta}^{\alpha} \qquad (\bm{\vartheta} \in S^{d-1}),
\end{equation}
where $\{ c_{\alpha} \}_{\alpha \in \mathbb{Z}_{+}^{d}}$ are the expansion coefficients.
By extending the domain of $\zeta$ to $\mathbb{R}^{d}$, we define
\begin{align}
    \acute{\zeta}(\bm{r}) &:= \sum_{\alpha \in \mathbb{Z}_{+}^{d}} \acute{c}_{\alpha} \bm{r}^{\alpha} \quad (\bm{r} \in \mathbb{R}^{d}), \label{eq:zeta_acute_pol} \\
    \acute{c}_{\alpha} &:= (- \mathrm{i}k)^{-|\alpha|} c_{\alpha}. \label{eq:c_acute_pol}
\end{align}
Then, the following equality holds (see Appendix~\ref{AppendixA}):
\begin{equation} \label{eq:relation_between_polynomial_derivation}
    \acute{\zeta} (\partial_{\bm{r}}) \mathrm{e}^{-\mathrm{i} k \bm{\vartheta} \cdot \bm{r}} = \zeta(\bm{\vartheta}) \mathrm{e}^{-\mathrm{i} k \bm{\vartheta} \cdot \bm{r}}.
\end{equation}
Using Eqs.~(\ref{eq:relation_between_polynomial_derivation}) and (\ref{eq:plane_wave_expansion}), Eq.~(\ref{eq:directional_input}) can be rewritten as
\begin{align} 
    \hat{s}(\bm{r}_{\mathrm{r}}) &= \int_{S^{d-1}} \tilde{P}_{\mathrm{b}}\left( k \bm{\vartheta} \right) \zeta(\bm{\vartheta}) \mathrm{e}^{-\mathrm{i} k \bm{\vartheta} \cdot \bm{r}_{\mathrm{r}}} \mathrm{d} \bm{\vartheta} \notag \\
    &= \acute{\zeta} (\partial_{\bm{r}_{\mathrm{r}}}) \int_{S^{d-1}} \tilde{P}_{\mathrm{b}}\left( k \bm{\vartheta} \right) \mathrm{e}^{-\mathrm{i} k \bm{\vartheta} \cdot \bm{r}_{\mathrm{r}}} \mathrm{d} \bm{\vartheta} \notag \\
    &= \acute{\zeta} (D) \hat{p}(\bm{r}_{\mathrm{r}}).
\end{align}
This expression implies that imparting directivity to the received signal can be regarded as applying a polynomial differential operator $\acute{\zeta}(D)$ to the sound field $\hat{p}$.
In other words, the process of introducing directivity corresponds to the action of a spatial differential operator on the sound field.

In the preceding discussion, the directional response was defined in terms of simple polynomial forms.
For instance, in practical numerical implementations, a directional response may be represented by a simple polynomial such as a dipole pattern oriented along the $i$-axis.
While this model is effective in such cases, Eq.~(\ref{eq:directivity_pol}) is generally a redundant representation.
This is because the dimension of the space of homogeneous polynomials of degree $\nu$, $\mathscr{P}_{\nu}$, is given by \cite{axler2013}
\begin{equation}
    \dim \mathscr{P}_{\nu} = \frac{(d+\nu-1)!}{\nu!(d-1)!}
\end{equation}
and since $L^{2}(S^{d-1}) = \bigoplus_{\nu=0}^{\infty} \mathscr{Y}_{\nu}$, the dimension of the space $\mathscr{Y}_{\nu}$ is 
\begin{equation} \label{eq:dim_Hk}
    \dim \mathscr{Y}_{\nu} = \frac{(d+2\nu-2)(d+\nu-3)!}{(d-2)!\nu!}.
\end{equation}
As an example, Table~\ref{table1} lists the dimensions of $\mathscr{P}_{\nu}$ and $\mathscr{Y}_{\nu}$ for each degree $\nu$ when $d=3$.
In general, the directional response can also be expressed as
\begin{equation} \label{eq:directivity_Ynm}
    \zeta(\bm{\vartheta}) = \sum_{\nu=0}^{\infty} \sum_{\mu=1}^{\mathrm{dim} \mathscr{Y}_{\nu}} c_{\nu}^{\mu} Y_{\nu}^{\mu}(\bm{\vartheta}).
\end{equation}
For instance, when $d=3$ and the expansion is truncated at fourth order, the number of functions used is nine—six fewer than in the polynomial representation.
Thus, the representation using simple polynomials is redundant, and in many cases, expressing the directional response in terms of spherical harmonics is more efficient.

\begin{table}[t]
    \caption{\label{table1} Dimensions of the spaces $\mathscr{P}_{\nu}$ and $\mathscr{Y}_{\nu}$ for $d=3$.}
    \centering
    \begin{tabular}{|@{\hspace{3mm}}c@{\hspace{3mm}}||@{\hspace{3mm}}c@{\hspace{3mm}}|@{\hspace{3mm}}c@{\hspace{3mm}}|}
        \hline
        $\nu$ & Dimension of $\mathscr{P}_{\nu}$ & dimension of $\mathscr{Y}_{\nu}$ \\
        \hline
        0 & 1 & 1 \\
        \hline
        1 & 3 & 3 \\
        \hline
        2 & 6 & 5 \\
        \hline
        3 & 10 & 7 \\
        \hline
        4 & 15 & 9 \\
        \hline
    \end{tabular}
\end{table}
Hence, the representation using spherical harmonics is more efficient in the sense that complex directional responses can be expressed using fewer basis functions.
In this case, using the bijective correspondence between the space $\mathscr{Y}_{\nu}$ of spherical harmonics of degree $\nu$ and the space $\mathscr{H}_{\nu}$ of homogeneous harmonic polynomials of the same degree $\nu$, we extend the domain as
\begin{equation} \label{eq:relation_Y_y}
    y_{\nu}^{\mu}(\bm{r}) = |\bm{r}|^{\nu} Y_{\nu}^{\mu} \left( \hat{\bm{r}} \right),
\end{equation}
where $y_{\nu}^{\mu}$ denotes the corresponding homogeneous harmonic polynomial.
Using this correspondence, Eqs.~(\ref{eq:zeta_acute_pol}) and (\ref{eq:c_acute_pol}) can be rewritten as
\begin{align}
    \acute{\zeta} (\bm{r}) &:= \sum_{\nu=0}^{\infty} \sum_{\mu=1}^{\mathrm{dim} \mathscr{Y}_{\nu}} \acute{c}_{\nu}^{\mu} y_{\nu}^{\mu}(\bm{r}), \label{eq:zeta_acute_Ynm} \\
    \acute{c}_{\nu}^{\mu} &:= (-\mathrm{i} k)^{-\nu} c_{\nu}^{\mu}. \label{eq:c_acute_Ynm}
\end{align}
In the following discussion, the directivity function is assumed to take the form of Eq.~(\ref{eq:directivity_Ynm}).
However, when it is expressed as in Eq.~(\ref{eq:directivity_pol}), the same discussion remains valid by appropriately converting the representation.

\subsection{\label{subsec:2.2} Sound Field Reconstruction}
In this subsection, we consider an internal source-free region $\Omega$ where directional sensors are placed at sampling points $\{ \bm{r}_{n} \}_{n=1}^{N}$.  
The receiving directivity function of the $n$-th sensor is assumed to be 
\begin{equation}
\zeta_{\mathrm{r}, n} = \sum_{\nu=0}^{\infty} \sum_{\mu=1}^{\mathrm{dim} \mathscr{Y}_{\nu}} c_{\nu}^{n, \mu} Y_{\nu}^{\mu}.
\end{equation}
Under the above conditions, the sound field belongs to the following reproducing kernel Hilbert space (RKHS) \cite{iwami2024jasa}:
\begin{align} \label{eq:def_S_k-d-1}
    &\mathcal{S}_{k}^{d-1} := \left\{ f \in L^{2}(\mathbb{R}^{d}) \mid \exists \tilde{F} \in L^{2}(k S^{d-1}) \right. \notag \\
    &\qquad \qquad \left. \mathrm{s.t.}\; f(\bm{r}) = \int_{S^{d-1}} \tilde{F} (k \bm{\vartheta}) \mathrm{e}^{\mathrm{i} k \bm{\vartheta} \cdot \bm{r}} \mathrm{d} \bm{\vartheta} \right\},
\end{align}
which is equipped with the inner product
\begin{equation}
    \langle f, g \rangle := \int_{S^{d-1}} \tilde{F} (k \bm{\vartheta}) \overline{\tilde{G} (k \bm{\vartheta})} \mathrm{d} \bm{\vartheta}.
\end{equation}
Its RK is given by
\begin{equation} \label{eq:RK_surface}
    \kappa_{k}(\bm{r}, \bm{r}^{\prime}) = \mathrm{J}_{d,0} (k|\bm{r}-\bm{r}^{\prime}|),
\end{equation}
where
\begin{equation} \label{eq:J_dn}
    \mathrm{J}_{d,\nu}(z) := (2\pi)^{\frac{d}{2}} \frac{J_{\nu+\frac{d}{2}-1}(z)}{z^{\frac{d}{2}-1}}
\end{equation}
\cite{iwami2025ast}, and $J_{\nu}$ denotes the Bessel function of order $\nu$ \cite{bowman1958}.
The RK can also be expressed in the following integral form:
\begin{equation} \label{eq:kappa_k_integral}
    \kappa_{k}(\bm{r}, \bm{r}^{\prime}) := \int_{S^{d-1}} \mathrm{e}^{\mathrm{i} k \bm{\vartheta} \cdot (\bm{r} - \bm{r}^{\prime})} \mathrm{d} \bm{\vartheta}.
\end{equation}

Now, considering the sound field reconstruction problem, the best approximation of the sound field can be represented using the RK as
\begin{equation} \label{eq:RK_model}
    \hat{p}_{\mathrm{est}} = \sum_{n=1}^{N} a_{n} \kappa_{k}(\cdot, \bm{r}_{n}).
\end{equation}
Using this representation, the output signal of the $n^{\prime}$-th directional element can be approximated as
\begin{align} 
    \hat{s}(\bm{r}_{n^{\prime}}) &= \acute{\zeta}_{\mathrm{r}, n^{\prime}}(D) \hat{p}(\bm{r}_{n^{\prime}}) \notag \\
    &\approx \sum_{n=1}^{N} a_{n} \sum_{\nu=0}^{\infty} \sum_{\mu=1}^{\mathrm{dim} \mathscr{Y}_{\nu}} \acute{c}_{\nu}^{n,\mu} y_{\nu}^{\mu}(\partial_{1}) \kappa_{k}(\bm{r}_{n^{\prime}}, \bm{r}_{n}).
\end{align}
When combined for all microphones, this can be expressed in matrix form as
\begin{equation}
    \hat{\bm{s}} = C \bm{a},
\end{equation}
where
\begin{align}
    \hat{\bm{s}} &:= 
    \begin{bmatrix}
        \hat{s}_{1}(\bm{r}_{1}) & \cdots & \hat{s}_{N}(\bm{r}_{N})
    \end{bmatrix}^{\mathsf{T}}, \\
    \bm{a} &:= 
    \begin{bmatrix}
        a_{1} & \cdots & a_{N}
    \end{bmatrix}^{\mathsf{T}}, \\
    C &= (C_{i,j}) := \left( \sum_{\nu=0}^{\infty} \sum_{\mu=1}^{\mathrm{dim} \mathscr{Y}_{\nu}} \acute{c}_{\nu}^{i,\mu} y_{\nu}^{\mu}(\partial_{1}) \kappa_{k}(\bm{r}_{i}, \bm{r}_{j}) \right). \label{eq:mat_D}
\end{align}
By solving this linear inverse problem, the coefficient vector $\bm{a}$ can be estimated.  
In a simple case where $C$ is nonsingular, it can be directly obtained as
\begin{equation} \label{eq:C_inv}
    \bm{a} = C^{-1} \hat{\bm{s}}.
\end{equation}
Depending on the sensor directivities or the number of elements, this inverse problem may become ill-posed.  
In such cases, the coefficients are estimated using Tikhonov regularization \cite{hansen2010}:
\begin{equation} \label{eq:C_Tikhonov}
    \bm{a} = (C^{\mathsf{T}} C + \lambda I)^{-1} C^{\mathsf{T}} \hat{\bm{s}},
\end{equation}
where $\lambda$ is the regularization parameter.

Next, we discuss the handling of spatial differentiation of the RK.  
As a preparation, we recall Hobson's theorem and its corollary \cite{nomura2018}.
\begin{prp}
(\textbf{Hobson's formula})\\
Assume that $f \in C^{\infty}(\mathbb{R}^{d})$ can be regarded as a radial function $f(\bm{r}) = f_{0}(|\bm{r}|)$.  
Then, for any $g \in \mathscr{P}_{\nu}$, the following relation holds:
\begin{equation} \label{eq:Hobson}
    g(\partial_{\bm{r}}) f(\bm{r}) = \sum_{j=0}^{\lfloor \nu/2 \rfloor} \frac{1}{2^{j}j!} 
    \left( \frac{1}{|\bm{r}|}\frac{\mathrm{d}}{\mathrm{d} |\bm{r}|} \right)^{\nu-j} 
    f_{0}(|\bm{r}|) \Delta^{j}g(\bm{r}).
\end{equation}
\end{prp}
From this, the following corollary immediately follows.
\begin{cor} \label{cor:Hobson_mod}
Assume that $f \in C^{\infty}(\mathbb{R}^{d})$ is a radial function $f(\bm{r}) = f_{0}(|\bm{r}|)$.  
Then, for any $h \in \mathscr{H}_{\nu}$, the following holds:
\begin{equation} \label{eq:Hobson_mod}
    h(\partial_{\bm{r}}) f(\bm{r}) = 
    \left( \frac{1}{|\bm{r}|}\frac{\mathrm{d}}{\mathrm{d} |\bm{r}|} \right)^{\nu} 
    f_{0} (|\bm{r}|) h(\bm{r}).
\end{equation}
\end{cor}
On the other hand, the following property of the derivative of the Bessel function is known \cite{martin2006}:
\begin{equation} \label{eq:0th-Bessel_differentiation}
    \left( \frac{1}{z}\frac{\mathrm{d}}{\mathrm{d} z} \right)^{\nu} 
    \left(\frac{1}{z^{n}} J_{n}(z) \right)
    = (-1)^{\nu}\frac{1}{z^{n+\nu}} J_{n+\nu}(z).
\end{equation}
In particular, when $n = 0$,
\begin{equation}
    \left( \frac{1}{z}\frac{\mathrm{d}}{\mathrm{d} z} \right)^{\nu}J_{0}(z)
    = (-1)^{\nu} \frac{1}{z^{\nu}} J_{\nu}(z).
\end{equation}

The radial functions appearing in the spherical harmonic representation of the interior problem can be uniformly expressed in general dimensions \cite{iwami2025ast,mclean2000}.  
Using this notation, the following relation holds for the special function $\mathrm{J}_{d,n}$ (see Appendix~\ref{AppendixB}):
\begin{equation}
    \left( \frac{1}{z}\frac{\mathrm{d}}{\mathrm{d} z} \right)^{\nu}\mathrm{J}_{d,0}(z)
    = (-1)^{\nu} \frac{1}{z^{\nu}} \mathrm{J}_{d,\nu}(z).
\end{equation}
Let $\kappa_{k,0}(|\bm{r}-\bm{r}^{\prime}|) := \kappa_{k}(\bm{r}, \bm{r}^{\prime})$, treating $\kappa_{k}$ as a single-variable function.  
Then, since $y_{\nu}^{\mu} \in \mathscr{H}_{\nu}$, the spatial derivative of the RK $y_{\nu}^{\mu}(\partial_{\bm{r}}) \kappa_{k}$ can be transformed using the above properties and the corollary as follows (with $\bm{s}:=\bm{r}-\bm{r}^{\prime}$):
\begin{align}
    y_{\nu}^{\mu}(\partial_{1}) \kappa_{k}(\bm{r}, \bm{r}^{\prime})
    &= \left( \frac{1}{|\bm{s}|} \frac{\mathrm{d}}{\mathrm{d} |\bm{s}|} \right)^{\nu} 
       \kappa_{k,0}(|\bm{s}|) y_{\nu}^{\mu}(\bm{s}) \notag \\
    &= \left( \frac{1}{|\bm{s}|} \frac{\mathrm{d}}{\mathrm{d} |\bm{s}|} \right)^{\nu} 
       \mathrm{J}_{d,0}(k |\bm{s}|) y_{\nu}^{\mu}(\bm{s}) \notag \\
    &= k^{2\nu} 
       \left( \frac{1}{k|\bm{s}|} \frac{\mathrm{d}}{\mathrm{d} (k|\bm{s}|)} \right)^{\nu} 
       \mathrm{J}_{d,0}(k |\bm{s}|) y_{\nu}^{\mu}(\bm{s}) \notag \\
    &= k^{2\nu} (-1)^{\nu} \frac{1}{k^{\nu} |\bm{s}|^{\nu}} 
       \mathrm{J}_{d,\nu}(k |\bm{s}|) |\bm{s}|^{\nu} 
       Y_{\nu}^{\mu}\!\left( \frac{\bm{s}}{|\bm{s}|} \right) \notag \\
    &= (-k)^{\nu} \mathrm{J}_{d,\nu}(k |\bm{r}-\bm{r}^{\prime}|) 
       Y_{\nu}^{\mu}\!\left( \frac{\bm{r}-\bm{r}^{\prime}}{|\bm{r}-\bm{r}^{\prime}|} \right).
       \label{eq:relation_RK_JY}
\end{align}
Therefore, by using the above equation and Eq.~(\ref{eq:c_acute_Ynm}), the $(i,j)$-th element of matrix $C$ can be expressed as
\begin{align}
    C_{i,j} 
    &= \sum_{\nu=0}^{\infty} \sum_{\mu=1}^{\mathrm{dim} \mathscr{Y}_{\nu}} 
       \acute{c}_{\nu}^{i,\mu} y_{\nu}^{\mu}(\partial_{1}) 
       \kappa_{k}(\bm{r}_{i}, \bm{r}_{j}) \notag \\
    &= \sum_{\nu=0}^{\infty} \sum_{\mu=1}^{\mathrm{dim} \mathscr{Y}_{\nu}} 
       \mathrm{i}^{-\nu} c_{\nu}^{i,\mu} 
       \mathrm{J}_{d,\nu}(k |\bm{r}_{i} - \bm{r}_{j}|) 
       Y_{\nu}^{\mu}\!\left( 
           \frac{\bm{r}_{i} - \bm{r}_{j}}{|\bm{r}_{i} - \bm{r}_{j}|}
       \right). \label{eq:Cij}
\end{align}
In numerical implementations, the series is truncated at a maximum order $\nu_{\mathrm{max}}$.  
This truncation physically corresponds to approximating the assumed sensor directivity function by a polynomial of order $\nu_{\mathrm{max}}$.

Consequently, sound field reconstruction becomes possible using microphone arrays with arbitrary directivities.  
In particular, when the directivity function is expressed as a spherical harmonic expansion, applying the induced spatial differential operator to the RK yields a linear combination of spherical basis functions for the interior problem.  
In this sense, the RK of the spherical surface band-limited function space $\mathcal{S}_{k}^{d-1}$ possesses remarkably favorable properties with respect to spatial differentiation.

\section{\label{sec:3} Beamformers in the RK Domain}
In this section, we formulate beamformers based on the sound-field representation using the RK.

\subsection{\label{subsec:3.1} Simple beamformer}
We first consider the simplest classical beamformer that steers a beam in a single direction.

Since we are dealing with an interior sound field, the model can be written as in Eq.~(\ref{eq:plane_wave_expansion}).  
To extract the input of a plane wave arriving from direction $\bm{\varphi}$ at position $\bm{r}$, define the linear functional $\rho_{\bm{\varphi}, \bm{r}} : \mathcal{S}_{k}^{d-1} \rightarrow \mathbb{C}$ by
\begin{equation} \label{eq:BF_functional_simple}
    \rho_{\bm{\varphi}, \bm{r}}(\hat{p}) = \tilde{P}_{\mathrm{b}} \left( k \bm{\varphi} \right) \mathrm{e}^{-\mathrm{i}k \bm{\varphi} \cdot \bm{r}}.
\end{equation}
By substituting Eq.~(\ref{eq:kappa_k_integral}) into the estimator of the field, Eq.~(\ref{eq:RK_model}), and rearranging, we obtain
\begin{equation}
    \hat{p}_{\mathrm{est}}(\bm{r}) = \int_{S^{d-1}} \left( \sum_{n=1}^{N} a_{n} \mathrm{e}^{\mathrm{i}k \bm{\vartheta} \cdot \bm{r}_{n}} \right) \mathrm{e}^{-\mathrm{i}k \bm{\vartheta} \cdot \bm{r}} \mathrm{d} \bm{\vartheta},
\end{equation}
which, by comparison with Eq.~(\ref{eq:plane_wave_expansion}), implies
\begin{equation} \label{eq:P_b_tilde_est}
    \tilde{P}_{\mathrm{b}}\left( k \bm{\vartheta} \right) \approx \sum_{n=1}^{N} a_{n} \mathrm{e}^{\mathrm{i}k \bm{\vartheta} \cdot \bm{r}_{n}}.
\end{equation}
Using this relation, Eq.~(\ref{eq:BF_functional_simple}) reduces to the simple form
\begin{align}
    \rho_{\bm{\varphi}, \bm{r}}(\hat{p}) &\approx \sum_{n=1}^{N} a_{n} \mathrm{e}^{\mathrm{i}k \bm{\varphi} \cdot (\bm{r}_{n} - \bm{r})} \notag \\
    &= \bm{w}_{\bm{\varphi}, \bm{r}}^{\mathsf{H}} \hat{\bm{s}}, \label{eq:simple_BF}
\end{align}
where $(\cdot)^{\mathsf{H}}$ denotes the Hermitian transpose.
The weight vector $\bm{w}_{\bm{\varphi}, \bm{r}}$ is defined by
\begin{equation} \label{eq:weight_simple}
    \bm{w}_{\bm{\varphi}, \bm{r}}^{\ast} := C^{-\mathsf{T}} 
    \begin{bmatrix}
        \mathrm{e}^{\mathrm{i}k \bm{\varphi} \cdot (\bm{r}_{1} - \bm{r})} & \cdots & \mathrm{e}^{\mathrm{i}k \bm{\varphi} \cdot (\bm{r}_{N} - \bm{r})}
    \end{bmatrix}^{\mathsf{T}},
\end{equation}
where $(\cdot)^{\ast}$ denotes the complex conjugate.
The above weight vector can be precomputed provided the microphone positions are known; thus, in practice the beamformer can be implemented simply by taking the (time-discrete) Fourier transform of the array inputs and computing an inner product.

\subsection{\label{subsec:3.2} General beamformer}
Next we consider beamformers that realize general beam shapes.  
Let the desired beam shape (directivity) function $\zeta \in L^{2}(S^{d-1})$ be assumed as
\begin{equation} \label{eq:desired_beam}
    \zeta(\bm{\vartheta}) = \sum_{\nu=0}^{\infty} \sum_{\mu=1}^{\mathrm{dim} \mathscr{Y}_{\nu}} c_{\nu}^{\mu} Y_{\nu}^{\mu}(\bm{\vartheta}).
\end{equation}
Using the associated spatial differential operator $\acute{\zeta}(D)$, the linear functional $\rho_{\zeta, \bm{r}} : \mathcal{S}_{k}^{d-1} \rightarrow \mathbb{C}$ that yields the desired input is expressed as
\begin{equation} \label{eq:bf}
    \rho_{\zeta, \bm{r}}(\hat{p}) = \acute{\zeta}(D)\hat{p}(\bm{r}).
\end{equation}
In other words, beamforming is regarded as a functional that returns the virtual directional input with directivity $\zeta$ at position $\bm{r}$.  
Note that if (formally, not as an $L^{2}$ function) one sets $\zeta(\bm{\theta}) = \delta(\bm{\theta} - \bm{\varphi})$, the simple beamformer above is recovered; hence the present formulation is a generalization of the previous subsection.

Since the sound field is approximated by Eq.~(\ref{eq:RK_model}), the beamformer output can be written as
\begin{align}
    \rho_{\zeta, \bm{r}}(\hat{p}) &\approx \sum_{n=1}^{N} a_{n} \acute{\zeta}(\partial_{1}) \kappa_{k}(\bm{r}, \bm{r}_{n}) \notag \\
    &= \bm{w}_{\zeta, \bm{r}}^{\mathsf{H}} \bm{p},
\end{align}
where the beamformer weight vector $\bm{w}_{\zeta, \bm{r}}$ is given by
\begin{equation} \label{eq:weight_complex}
    \bm{w}_{\zeta, \bm{r}}^{\ast} := C^{-\mathsf{T}} 
    \begin{bmatrix}
        \acute{\zeta}(\partial_{1}) \kappa_{k}(\bm{r}, \bm{r}_{1}) & \cdots & \acute{\zeta}(\partial_{1}) \kappa_{k}(\bm{r}, \bm{r}_{N})
    \end{bmatrix}^{\mathsf{T}}.
\end{equation}
This weight vector can also be precomputed, so real-time processing reduces to inner-product operations on the frequency-domain array inputs.

The spatial derivative of the RK is obtained from Eq.~(\ref{eq:desired_beam}) and the results of the previous section as
\begin{equation} \label{eq:derivative_RK}
    \acute{\zeta}(\partial_{1}) \kappa_{k}(\bm{r}, \bm{r}^{\prime}) 
    = \sum_{\nu=0}^{\infty} \sum_{\mu=1}^{\mathrm{dim} \mathscr{Y}_{\nu}} \mathrm{i}^{-\nu} c_{\nu}^{\mu} \mathrm{J}_{d,\nu}(k |\bm{r} - \bm{r}^{\prime}|) 
    Y_{\nu}^{\mu}\!\left( \frac{\bm{r} - \bm{r}^{\prime}}{|\bm{r} - \bm{r}^{\prime}|} \right).
\end{equation}

This beamformer therefore offers high degrees of freedom: it can acquire inputs for arbitrary directivities (not limited to axisymmetric patterns) and at arbitrary positions.

\subsection{\label{subsec3.3} Extraction of Directional Sound Fields by Beamforming}
Many existing beamformers implicitly assume a directional input located at the origin.  
In contrast, the proposed method allows explicit specification of the spatial location from which to extract the signal.  
Exploiting this fact, whereas conventional beamformers vary $\bm{\varphi}$ or $\zeta$ while keeping $\bm{r}=\bm{0}$ in Eqs.~(\ref{eq:weight_simple}) and (\ref{eq:weight_complex}), here we fix $\bm{\varphi}$ or $\zeta$ and vary $\bm{r}$ to extract the directional sound field.

Based on Eqs.~(\ref{eq:weight_simple}) and (\ref{eq:weight_complex}), define the matrix that stacks the weight vectors for multiple desired points $\{ \bm{r}_{m}^{\prime} \}_{m=1}^{M}$ as
\begin{equation}
    W_{\zeta} := \begin{bmatrix}
        \bm{w}_{\zeta, \bm{r}_{1}^{\prime}} & \cdots & \bm{w}_{\zeta, \bm{r}_{M}^{\prime}}
    \end{bmatrix}.
\end{equation}
Then, the extracted directional field corresponding to the desired beam is obtained by
\begin{equation}
    \bm{p}_{\zeta} = W_{\zeta}^{\mathsf{H}} \hat{\bm{s}}.
\end{equation}

This operation is equivalent to observing the field after applying the directional weight $\zeta$ to the incoming directions.  
Therefore, it can be used to emphasize, extract, or suppress components arriving from particular directions and to visualize the resulting directional sound field.

\section{\label{sec:4} Interpretation of Axisymmetric Beamformers in the Spherical Harmonic Domain}
In the previous section, beamforming was formulated as a functional realized by spatial differentiation.  
By introducing the same concept, the axisymmetric beamformers \cite{meyer2002,boaz2015} can be reinterpreted within this framework, and it is further shown that non-axisymmetric beamformers can also be implemented in the spherical harmonic domain.

\subsection{\label{subsec:4.1} Axisymmetric Beamformers in the Spherical Harmonic Domain}
We here review the framework of conventional methods.  
The spatial dimension is taken as $d=3$.

The interior sound field is expanded using spherical harmonics \cite{williams1999}:
\begin{equation} \label{eq:SH_expansion}
    \hat{p}(\bm{r}) = \sum_{\nu=1}^{\infty} \sum_{\mu = -\nu}^{\nu} \hat{p}_{\nu}^{\mu} \, \mathrm{J}_{3, \nu}(k|\bm{r}|) \, Y_{\nu}^{\mu}\!\left( \frac{\bm{r}}{|\bm{r}|} \right),
\end{equation}
where $\hat{p}_{\nu}^{\mu}$ are the expansion coefficients of the field.  
In the context of spherical arrays, the signals measured by rigid-sphere microphones or directional microphones on a radius $a$ can be written using radial functions $b_{\nu}$ that do not have zeros as
\begin{equation}
    \hat{s}(\bm{r}) = \sum_{\nu=1}^{\infty} \sum_{\mu = -\nu}^{\nu} \hat{p}_{\nu}^{\mu} \, b_{\nu}(ka) \, Y_{\nu}^{\mu}\!\left( \frac{\bm{r}}{a} \right),
\end{equation}
(note that $|\bm{r}| = a$).  
In what follows we work with a truncated approximation up to order $\nu_{\mathrm{max}}$.

Define the weight vector in the spherical harmonic domain with components
\begin{equation} \label{eq:conventional_weigh_vec}
    w_{\nu}^{\mu} = \frac{\mathrm{i}^{\nu} d_{\nu}}{b_{\nu}(ka)} \, Y_{\nu}^{\mu}(\bm{\varphi}),
\end{equation}
where $\bm{\varphi}$ denotes the look direction (the main-lobe direction) and $d_{\nu}$ are the beamforming weights.  
Using these weights, an axisymmetric beamformer is realized by
\begin{align}
    y &= \bm{w}^{\mathsf{H}} \bm{s}, \\
    \bm{w}^{\ast} &:= S^{\mathsf{T}} \bm{w}_{\nu}^{\mu},
\end{align}
where $S$ denotes the discrete spherical Fourier transform matrix.  
To see the beam pattern this produces, compute the beamformer output for an incident plane wave from direction $\bm{\vartheta}$ (for which $\hat{p}_{\nu}^{\mu} = \mathrm{i}^{-\nu} \overline{Y_{\nu}^{\mu}(\bm{\vartheta})}$):
\begin{align}
    y &\approx \sum_{\nu=1}^{\nu_{\mathrm{max}}} \sum_{\mu = -\nu}^{\nu} \frac{\mathrm{i}^{\nu} d_{\nu}}{b_{\nu}(ka)} \, Y_{\nu}^{\mu}(\bm{\varphi}) \, \mathrm{i}^{-\nu} b_{\nu}(ka) \, \overline{Y_{\nu}^{\mu}(\bm{\vartheta})} \notag \\
    &= \sum_{\nu=1}^{\nu_{\mathrm{max}}} d_{\nu} \sum_{\mu = -\nu}^{\nu} Y_{\nu}^{\mu}(\bm{\varphi}) \overline{Y_{\nu}^{\mu}(\bm{\vartheta})} \notag \\
    &= \sum_{\nu=1}^{\nu_{\mathrm{max}}} d_{\nu} \frac{2\nu + 1}{4\pi} \, P_{\nu}(\cos \Theta),
\end{align}
where in the last equality we used the addition theorem for spherical harmonics \cite{boaz2015,martin2006}.  
Here $\Theta$ is the angle between $\bm{\varphi}$ and $\bm{\vartheta}$, and $P_{\nu}$ denotes the Legendre polynomial of degree $\nu$ (equivalent to the order-0 spherical harmonic of degree $\nu$).

\subsection{\label{subsec:4.2} Interpretation via Spatial Differential Operators}
We now generalize the discussion to arbitrary spatial dimensions.  
Within the framework that regards spatial differential operators as a means to impose directivity, we first restrict Eq.~(\ref{eq:desired_beam}) to axisymmetric cases and rewrite it as
\begin{equation}
    \zeta(\bm{\vartheta}) = \sum_{\nu=0}^{\infty} d_{\nu} \, \Psi_{\nu}(\bm{\vartheta}, \bm{e}_{d}),
\end{equation}
where the real-valued function $\Psi_{\nu}: S^{d-1} \times S^{d-1} \rightarrow \mathbb{R}$ is the RK of $\mathscr{Y}_{\nu}$, and $\Psi_{\nu}(\cdot, \bm{e}_{d})$ is proportional to a Gegenbauer polynomial \cite{andrews1998}.  
Using a rotation $\ell \in SO(d,\mathbb{R})$ that steers the beam toward the look direction $\bm{\varphi}$, the directivity function can be written as
\begin{equation} \label{eq:axisymmetric_beam}
    \zeta(\bm{\vartheta}) = \sum_{\nu=0}^{\infty} d_{\nu} \, \Psi_{\nu}(\ell \bm{\vartheta}, \bm{e}_{d}).
\end{equation}
Since $\{ Y_{\nu}^{\mu} \}_{\mu=-\nu}^{\nu}$ is an orthonormal basis of $\mathscr{Y}_{\nu}$, the RK property gives
\begin{equation}
    \Psi_{\nu}(\bm{\vartheta}, \bm{\varphi}) = \sum_{\mu=1}^{\mathrm{dim} \mathscr{Y}_{\nu}} Y_{\nu}^{\mu}(\bm{\vartheta}) \overline{Y_{\nu}^{\mu}(\bm{\varphi})}.
\end{equation}
Using the $SO(d,\mathbb{R})$-invariance of the kernel $\Psi_{\nu}(\ell \bm{\vartheta}, \ell \bm{\varphi}) = \Psi_{\nu}(\bm{\vartheta}, \bm{\varphi})$ and its Hermitian symmetry $\Psi_{\nu}(\bm{\varphi}, \bm{\vartheta}) = \overline{\Psi_{\nu}(\bm{\vartheta}, \bm{\varphi})}$, Eq.~(\ref{eq:axisymmetric_beam}) can be rewritten as
\begin{equation}
    \zeta(\bm{\vartheta}) = \sum_{\nu=0}^{\infty} \sum_{\mu=1}^{\mathrm{dim} \mathscr{Y}_{\nu}} d_{\nu} \, Y_{\nu}^{\mu}(\ell^{-1} \bm{e}_{d}) \, \overline{Y_{\nu}^{\mu}(\bm{\vartheta})}.
\end{equation}

Now, for interior fields (\ref{eq:SH_expansion}), when the differential operator induced by $\overline{Y_{\nu}^{\mu}}$ acts on the field, the following relation holds (see Appendix~\ref{AppendixC}):
\begin{equation}
    \overline{y_{\nu}^{\mu}}(D) \, \hat{p}(\bm{0}) = k^{\nu} \, \hat{p}_{\nu}^{\mu}.
\end{equation}
Hence, the beamforming functional at the origin corresponding to the beam in Eq.~(\ref{eq:axisymmetric_beam}) is given by
\begin{align}
    \rho_{\zeta, \bm{0}}(\hat{p})
    &= \sum_{\nu=0}^{\nu_{\mathrm{max}}} \sum_{\mu=1}^{\mathrm{dim} \mathscr{Y}_{\nu}} \acute{d}_{\nu} \, Y_{\nu}^{\mu}(\ell^{-1} \bm{e}_{d}) \, \overline{y_{\nu}^{\mu}}(D) \, \hat{p}(\bm{0}) \notag \\
    &= \sum_{\nu=0}^{\nu_{\mathrm{max}}} \sum_{\mu=1}^{\mathrm{dim} \mathscr{Y}_{\nu}} \mathrm{i}^{\nu} \, \acute{d}_{\nu} \, Y_{\nu}^{\mu}(\ell^{-1} \bm{e}_{d}) \, \hat{p}_{\nu}^{\mu}.
\end{align}
Therefore, the corresponding weight vector in the spherical harmonic domain is
\begin{equation}
    w_{\nu}^{\mu} = \frac{\mathrm{i}^{\nu} d_{\nu}}{b_{\nu}(ka)} \, Y_{\nu}^{\mu}(\ell^{-1} \bm{e}_{d}),
\end{equation}
which coincides with Eq.~(\ref{eq:conventional_weigh_vec}).

In other words, conventional axisymmetric beamformers assume an axisymmetric beam and correspond to acquiring the input at the origin.
Conversely, if one wishes to extend the directivity function to non-axisymmetric patterns and arbitrary positions, assume
\begin{equation}
    \zeta(\bm{\vartheta}) = \sum_{\nu=0}^{\infty} \sum_{\mu=1}^{\mathrm{dim} \mathscr{Y}_{\nu}} c_{\nu}^{\mu} \, \overline{Y_{\nu}^{\mu}(\bm{\vartheta})},
\end{equation}
then one can compute
\begin{align}
    \rho_{\zeta, \bm{r}}(\hat{p})
    &= \sum_{\nu=0}^{\nu_{\mathrm{max}}} \sum_{\mu=1}^{\mathrm{dim} \mathscr{Y}_{\nu}} \acute{c}_{\nu}^{\mu} \, \overline{y_{\nu}^{\mu}}(D) \, \hat{p}(\bm{0}) \notag \\
    &= \sum_{\nu=0}^{\nu_{\mathrm{max}}} \sum_{\mu=1}^{\mathrm{dim} \mathscr{Y}_{\nu}} \mathrm{i}^{\nu} \, c_{\nu}^{\mu} \, \hat{p}_{\nu}^{\mu},
\end{align}
so that the corresponding spherical-harmonic-domain weight vector becomes
\begin{equation}
    w_{\nu}^{\mu} = \frac{\mathrm{i}^{\nu} c_{\nu}^{\mu}}{b_{\nu}(ka)}.
\end{equation}
Beamformers defined in other domains employing different basis functions can also be considered within the same framework.

\section{\label{sec:5}Numerical Simulations}
To verify the performance of the proposed method, three numerical simulations were conducted under the assumption of a two-dimensional sound field. 
The first examines sound field reconstruction, the second investigates beamforming performance, and the third evaluates directional field extraction.

\begin{figure}[t]
    \centering
    \includegraphics[width=1.\columnwidth]{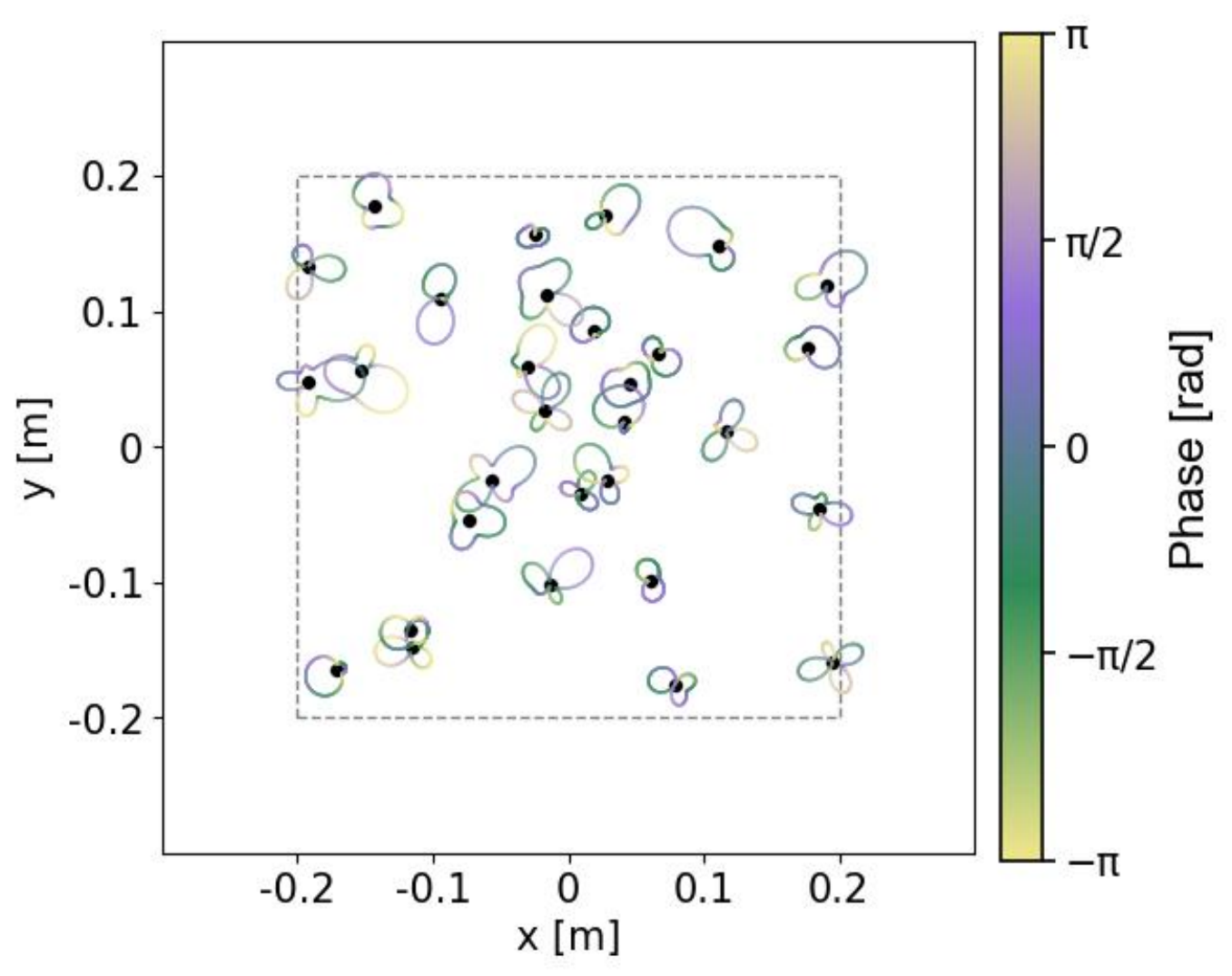} \vspace{0pt}
    \caption{\label{fig1}(Color online) Microphone configuration used in the numerical simulations.}
    \vspace{0pt}
\end{figure}

\begin{table}[t]
    \caption{\label{table2}Common parameters used in all numerical simulations.}
    \centering
    \begin{tabular}{l@{\hspace{15mm}}l}
        \hline\hline
        Maximum order of directivities ($\nu_{\mathrm{max}}$) & 2 \\
        Side length of microphone region & 0.4~m\\
        Sampling points $N$ & 30~samples \\
        Signal-to-noise ratio & 30~dB\\
        Regularization parameter & 0.001 \\
        \hline\hline
    \end{tabular}
\end{table}
The parameters common to all simulations are summarized in Table~\ref{table2}.
Thirty microphones were randomly distributed within a square region of 0.4~m on each side.
Each microphone had a maximum directivity order of two, and its coefficients were assigned randomly.
The microphone configuration is illustrated in Fig.~\ref{fig1}, where black dots denote the microphone positions and the local polar plots around each dot represent the directivity patterns. 
The distance from each microphone indicates sensitivity, and color denotes phase.  
For practical relevance, the signal-to-noise ratio (SNR) was set to 30~dB, and the regularization parameter for the coefficient estimation was fixed at 0.001.

\subsection{\label{subsec:5.1} Sound Field Reconstruction Performance}
High-quality beamforming requires accurate estimation of the sound field expansion coefficients, i.e., precise sound field reconstruction.
Under the conditions shown in Fig.~\ref{fig1}, a plane wave arriving from $45^{\circ}$ was reconstructed.
The performance was quantified using the mean normalized error (MNE) defined as
\begin{equation}
    \mathrm{MNE} = \frac{1}{M} \sum_{m=1}^{M} 20 \log_{10} \frac{|\hat{p}(\bm{r}_{m}^{\prime})-\hat{p}_{\mathrm{est}}(\bm{r}_{m}^{\prime})|}{|\hat{p}(\bm{r}_{m}^{\prime})|},
\end{equation}
where $M$ is the number of evaluation points and $\bm{r}_{m}^{\prime}$ denotes the $m$-th evaluation position.
The evaluation region was a square area of 0.5~m on each side, and the results are expressed in dB.

\begin{figure}[t]
    \centering
    \includegraphics[width=1.\columnwidth]{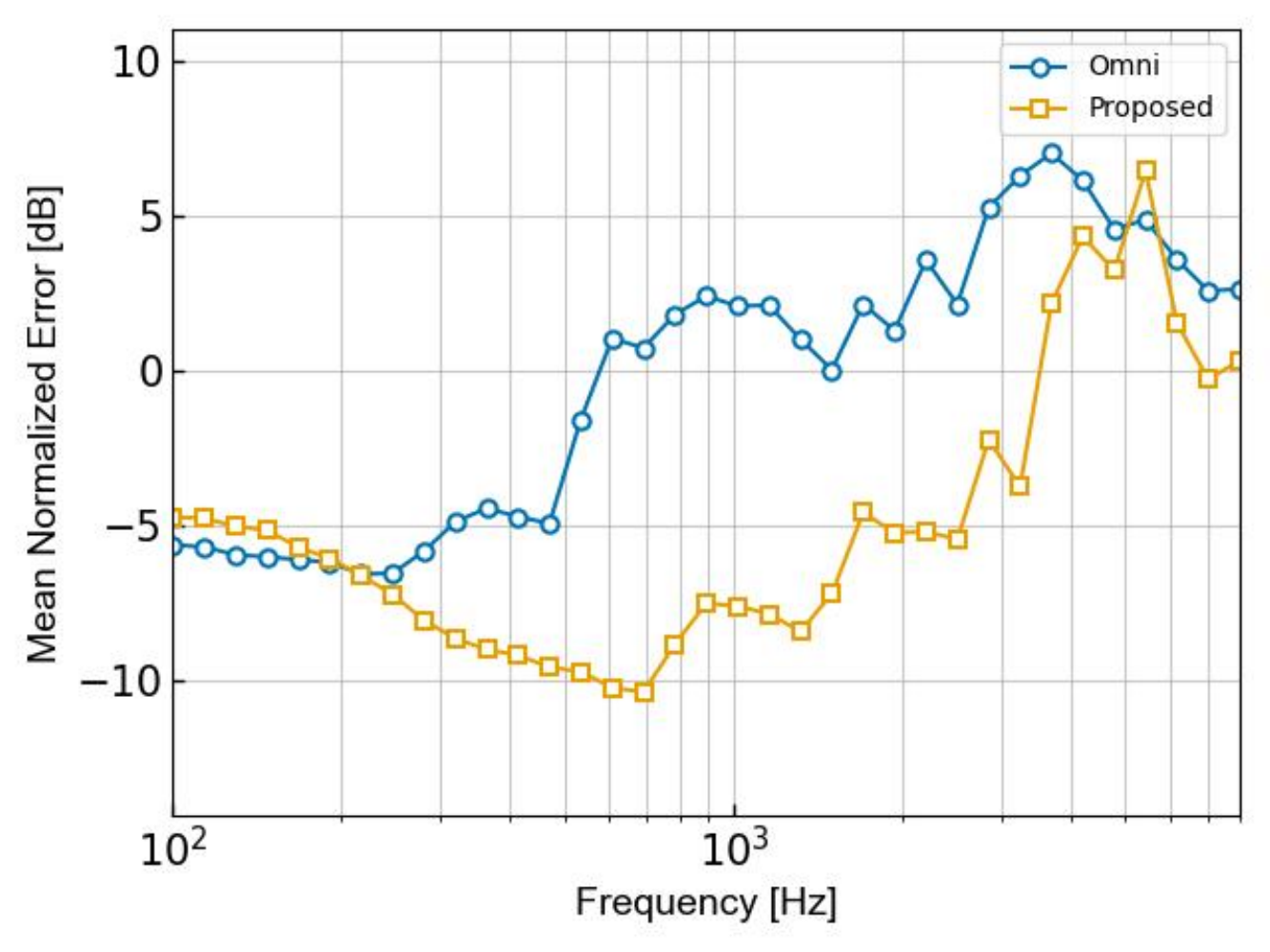} \vspace{0pt}
    \caption{\label{fig2}(Color online) MNEs of the existing and proposed methods.}
    \vspace{0pt}
\end{figure}

The experimental results are shown in Fig.~\ref{fig2}.  
To evaluate the effect of explicitly modeling microphone directivity, the case assuming omnidirectional microphones is labeled \textbf{Omni}, and the proposed method is labeled \textbf{Proposed}.

\begin{figure*}[t]
    \centering
    \includegraphics[width=1.6\columnwidth]{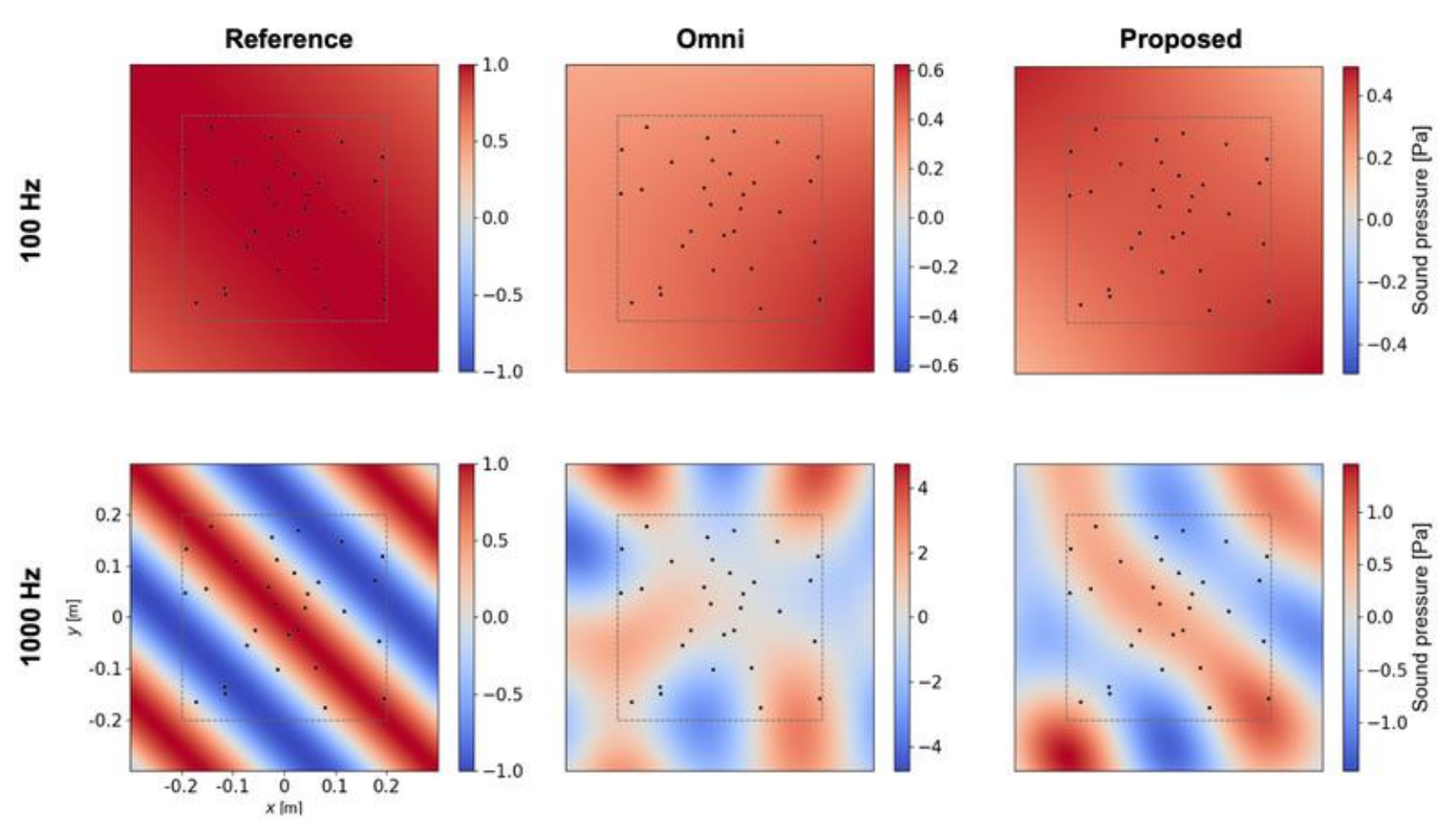}
    \caption{\label{fig3}(Color online) Reference and reconstructed sound fields at 100~Hz and 1000~Hz.}
    \vspace{0pt}
\end{figure*}

As shown in the figure, the proposed method exhibits smaller errors over a wide frequency range except for very low and very high frequencies.
At high frequencies, both methods are affected by spatial aliasing determined by the microphone geometry.
Examples of reconstructed sound fields at 100~Hz and 1000~Hz are shown in Fig.~\ref{fig3}.
At 1000~Hz, where the difference in MNE is large, the proposed method successfully reconstructs the plane-wave field in the microphone region, whereas the Omni method fails to do so.
This can be attributed to the fact that directional responses include phase variations (i.e., delays), and ignoring them yields almost no directional information.
At 100~Hz, where the MNEs of both methods are nearly identical, the results appear similar because the directional effect is negligible in the low-frequency range—the term $\mathrm{J}_{d,\nu}(k|\bm{r}_{i} - \bm{r}_{j}|)$ in Eq.~(\ref{eq:Cij}) becomes small, and the omnidirectional component dominates.
These results demonstrate that the proposed coefficient estimation considering microphone directivity is effective below the spatial aliasing frequency and in the frequency range where directional components significantly contribute.

\subsection{\label{subsec:5.2} Beamforming Performance}
To evaluate basic beamforming performance, the directivity index (DI) \cite{boaz2015,van2002} was calculated for each frequency.
The beamformer was implemented according to Eq.~(\ref{eq:simple_BF}) with the look direction $\bm{\varphi}$ at the origin.
The DI is defined as
\begin{equation} \label{eq:def_DI}
    \mathrm{DI} = 10 \log_{10} \frac{|S^{d-1}| |y(\bm{\varphi})|^{2} }{\int_{S^{d-1}} |y(\bm{\vartheta})|^{2} \mathrm{d} \bm{\vartheta} },
\end{equation}
where $y$ is defined for each direction $\bm{\vartheta} \in S^{d-1}$ as $y(\bm{\vartheta}) := \rho_{\bm{\varphi}, \bm{0}}(\hat{p}_{\bm{\vartheta}})$, and $\hat{p}_{\bm{\vartheta}}(\bm{r}) = \exp{(-\mathrm{i} k \bm{\vartheta} \cdot \bm{r})}$ is a plane wave arriving from direction $\bm{\vartheta}$.
Using Eqs.~(\ref{eq:simple_BF}) and (\ref{eq:kappa_k_integral}), Eq.~(\ref{eq:def_DI}) can be rewritten as
\begin{align}
    \mathrm{DI} &= 10 \log_{10} \frac{\bm{w}^{\mathsf{H}} \mathcal{S} \bm{w}}{\bm{a}^{\mathsf{H}} \mathcal{K} \bm{a}}, \notag \\
    \mathcal{S} &= |S^{d-1}| \bm{s} \bm{s}^{\mathsf{H}}, \notag \\
    \mathcal{K} &= (\kappa_{k}(\bm{r}_{i}, \bm{r}_{j})).
\end{align}
Here, $\mathcal{K}$ is the Gram matrix.

\begin{figure}[t]
    \centering
    \includegraphics[width=1.\columnwidth]{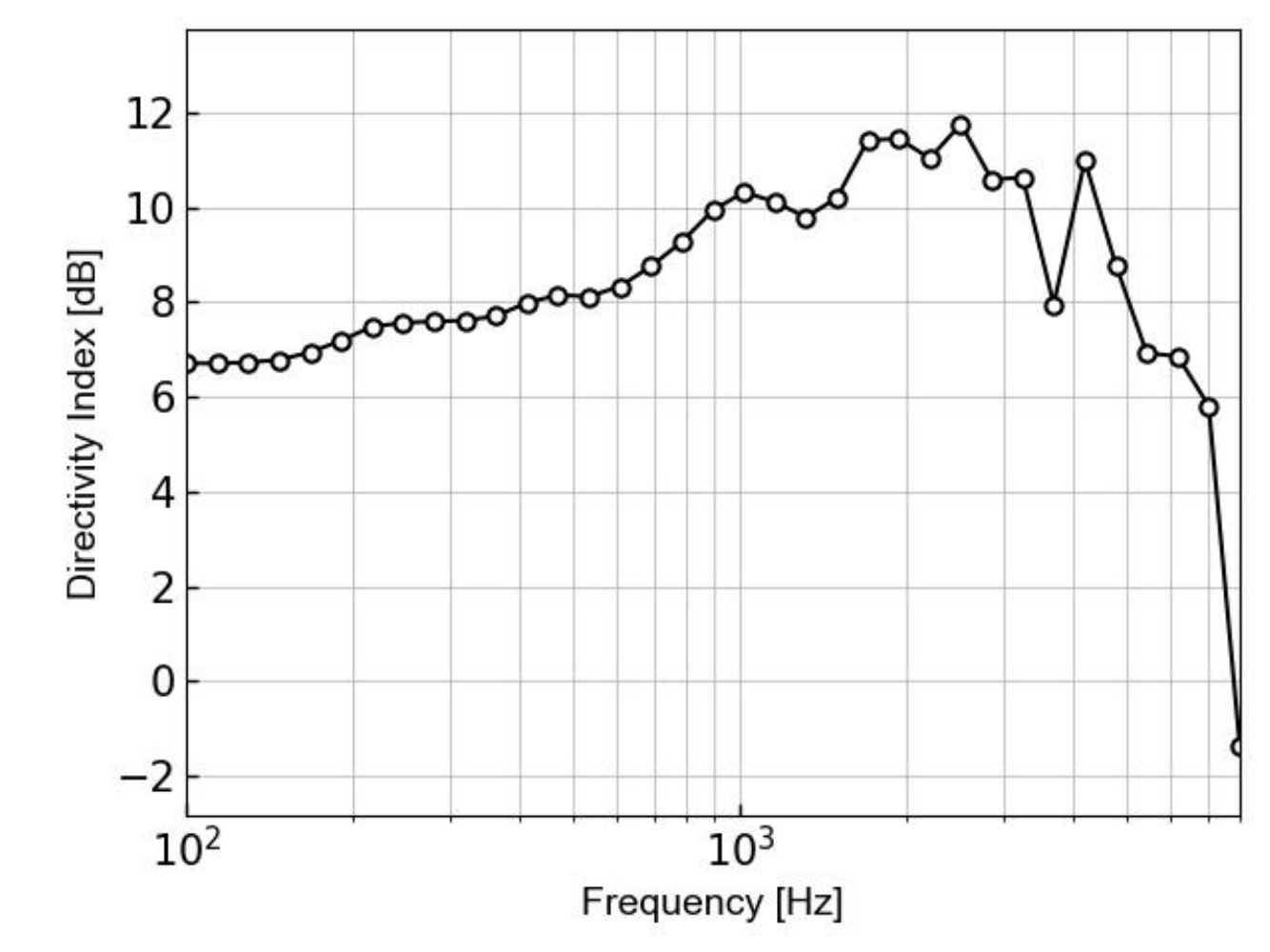} \vspace{0pt}
    \caption{\label{fig4}(Color online) Directivity indices (DIs) of the proposed method.}
    \vspace{0pt}
\end{figure}
\begin{figure*}[t]
    \centering
    \includegraphics[width=1.4\columnwidth]{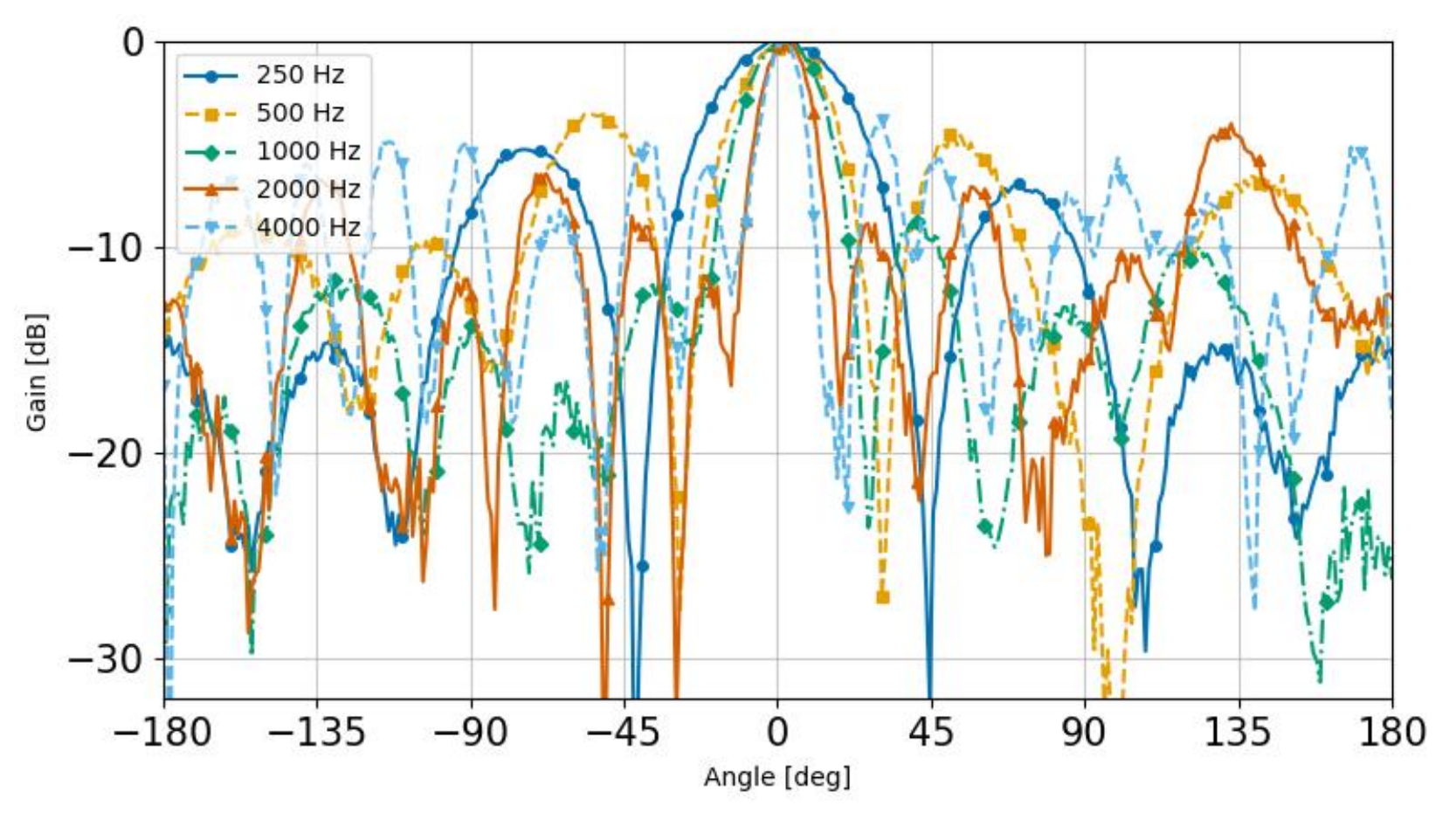}
    \caption{\label{fig5}(Color online) Beam patterns of the proposed method at 250, 500, 1000, 2000, and 4000~Hz.}
    \vspace{0pt}
\end{figure*}

The calculated DIs are shown in Fig.~\ref{fig4}, and the corresponding beam patterns at 250, 500, 1000, 2000, and 4000~Hz are presented in Fig.~\ref{fig5}.
The small oscillations in the gain curves are caused by the added noise.
Up to approximately 2~kHz, the DI increases monotonically, indicating sharper main lobes and smaller side lobes with increasing frequency.
Beyond 4~kHz, however, the DI drops sharply and the beam patterns become distorted, which can be attributed to spatial aliasing effects similar to those observed in the previous subsection.

\subsection{\label{subsec:5.3} Extraction of Directional Fields}
Finally, the performance of the directional field extraction algorithm described in Subsec.~\ref{subsec3.3} was evaluated.
A simple beamformer with look direction $\bm{\varphi} = [\cos(\pi/4), \sin(\pi/4)]$ at the origin was used, as in the previous subsections.

To simulate a sound field containing components arriving from multiple directions, the primary field was set to $\kappa_{k}(\bm{0}, \cdot)$, since the integral form of the RK (\ref{eq:kappa_k_integral}) represents a superposition of plane waves of equal amplitude and phase from all directions.
In this case, the desired field to be extracted is $\hat{p}_{\mathrm{des}}(\bm{r}) = \exp{(-\mathrm{i}k \bm{\varphi} \cdot \bm{r})}$.
The microphone input signals were analytically computed using Eq.~(\ref{eq:derivative_RK}).
The MNE with respect to the desired field was used as the evaluation metric.

\begin{figure}[t]
    \centering
    \includegraphics[width=1.\columnwidth]{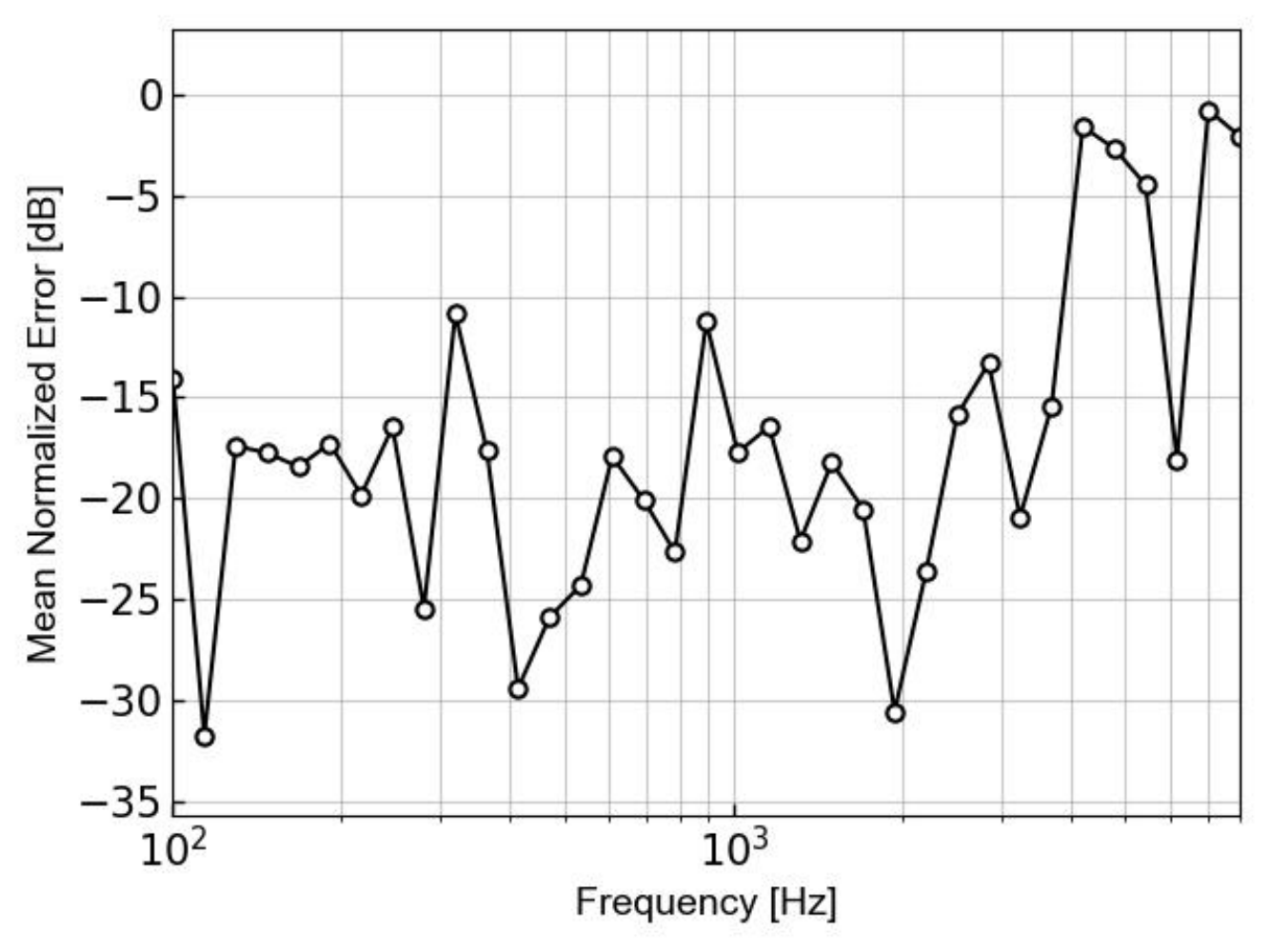} \vspace{0pt}
    \caption{\label{fig6}(Color online) MNEs of the extracted fields.}
    \vspace{0pt}
\end{figure}
\begin{figure}[t]
    \centering
    \includegraphics[width=1.\columnwidth]{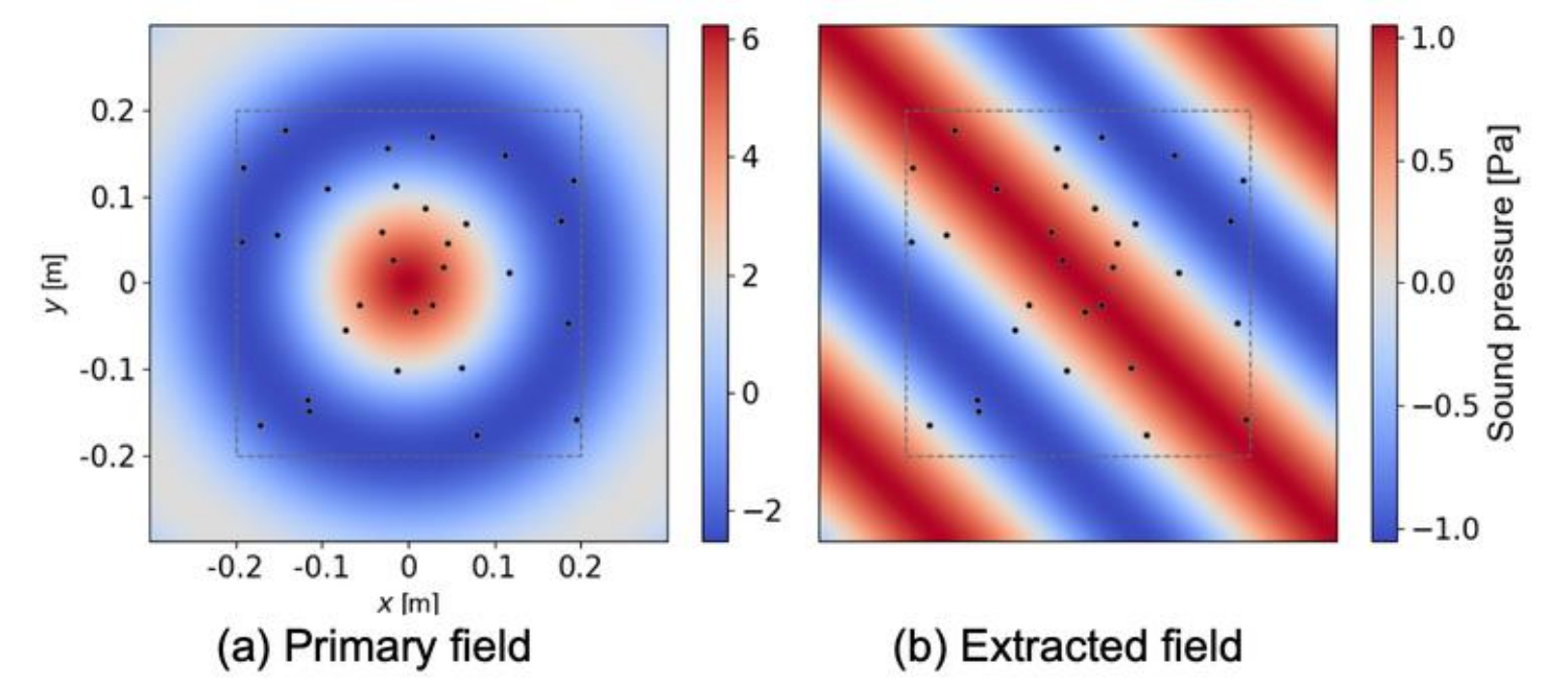} \vspace{0pt}
    \caption{\label{fig7}(Color online) (a) Reference field and (b) extracted field at 1000~Hz.}
    \vspace{0pt}
\end{figure}

The results are shown in Fig.~\ref{fig6}, and an example of the primary and extracted fields at 1000~Hz is presented in Fig.~\ref{fig7}.
The overall MNE trend is similar to that in Subsection~\ref{subsec:5.1}, approaching 0~dB at high frequencies.
However, the MNE values are generally smaller because, as indicated by Eq.~(\ref{eq:simple_BF}), the extracted field is of the form 
$A \exp{(-\mathrm{i} (k \bm{\varphi} \cdot \bm{r}) + \theta_{0})}$, 
reducing the problem to the simple estimation of amplitude and phase.

\begin{figure}[t]
    \centering
    \includegraphics[width=1.\columnwidth]{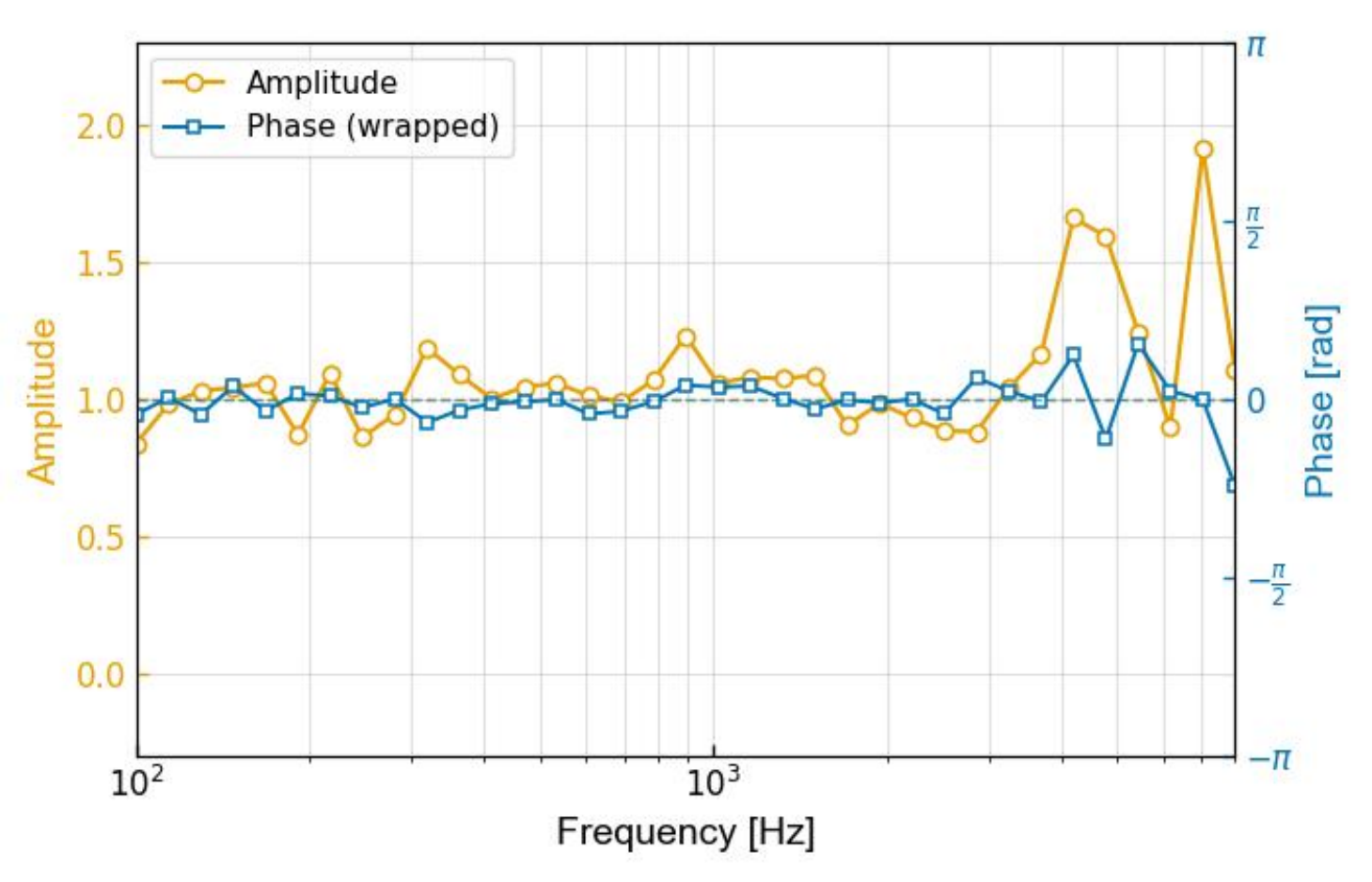} \vspace{0pt}
    \caption{\label{fig8}(Color online) Estimated amplitude and phase.}
    \vspace{0pt}
\end{figure}

Figure~\ref{fig8} shows the estimated amplitude and phase, obtained as the magnitude and argument of Eq.~(\ref{eq:P_b_tilde_est}).  
It can be seen that both quantities are accurately estimated over a wide frequency range except for the high-frequency region, which explains the reduction in MNE.

\section{\label{sec:6} Conclusion}
In this paper, we have systematically reformulated directional reception in arbitrary dimensions as spatial differentiation using polynomial differential operators and, based on this formulation, proposed a beamformer in the RK domain.
This formulation admits a simple development because the spatial differentiation of the RK reduces to a compact form via Hobson’s formula.
The proposed method is also applicable to the extraction of directional sound fields.
Furthermore, we have reinterpreted conventional axisymmetric beamformers in the spherical harmonic domain in terms of spatial differential operators and demonstrated their extensibility to non-axisymmetric beamformers.
The effectiveness of the proposed approach has been confirmed through three numerical simulations.

\begin{acknowledgments}
This research was partially supported by JSPS KAKENHI under Grant Number JP24K03222.
\end{acknowledgments}

\section*{AUTHOR DECLARATIONS}
\subsection*{Conflict of Interest}
The authors have no conflict of interest.

\section*{DATA AVAILABILITY}
The data that support the findings of this study are available from the corresponding author upon reasonable request.

\appendix
\section{\label{AppendixA} Proof that multiplication by a function on a plane wave is equivalent to applying a differential operator}
Let $f(\bm{r}) := \exp(-\mathrm{i} k \bm{\vartheta}\cdot \bm{r})$. Then
\begin{align} 
    D^{\alpha} f(\bm{r}) &= \frac{\partial^{|\alpha|}}{\partial x_{1}^{\alpha_{1}} \cdots \partial x_{d}^{\alpha_{d}}} \mathrm{e}^{-\mathrm{i} k \bm{\vartheta}\cdot \bm{r}} \notag \\
    &= (-ik)^{|\alpha|} \bm{\vartheta}^{\alpha} \mathrm{e}^{-\mathrm{i} k \bm{\vartheta}\cdot \bm{r}}.
\end{align}
Hence,
\begin{align} 
    \acute{\zeta}_{\mathrm{r}} (D) f(\bm{r}) &= \sum_{\alpha \in \mathbb{Z}_{+}^{d}} \acute{c}_{\mathrm{r}, \alpha} \partial_{\bm{r}}^{\alpha} \mathrm{e}^{-\mathrm{i} k \bm{\vartheta} \cdot \bm{r}} \notag \\
    &= \sum_{\alpha \in \mathbb{Z}_{+}^{d}} (-\mathrm{i} k)^{-|\alpha|} c_{\mathrm{r}, \alpha} (-\mathrm{i} k)^{|\alpha|} \mathrm{e}^{-\mathrm{i} k \bm{\vartheta}\cdot \bm{r}} \notag \\
    &= \zeta_{\mathrm{r}}(\bm{\vartheta}) \mathrm{e}^{-\mathrm{i} k \bm{\vartheta} \cdot \bm{r}}
\end{align}

\section{\label{AppendixB} Proof of the differentiation property of $\mathrm{J}_{d,n}$}
The claim follows from the following algebraic manipulations:
\begin{align}
    \mathrm{J}_{d,\nu}(z) &= (2\pi)^{\frac{d}{2}} \frac{1}{z^{\frac{d}{2}-1}} J_{\nu+\frac{d}{2}-1}(z) \notag \\
    &= (2\pi)^{\frac{d}{2}} z^{\nu} \frac{1}{z^{\nu+\frac{d}{2}-1}} J_{\nu+\frac{d}{2}-1}(z) \notag \\
    &= (-1)^{\nu} z^{\nu} \left( \frac{1}{z}\frac{\mathrm{d}}{\mathrm{d} z} \right)^{\nu} \left( (2\pi)^{\frac{d}{2}} \frac{1}{z^{\frac{d}{2}-1}} J_{\frac{d}{2}-1}(z) \right) \notag \\
    &= (-1)^{\nu} z^{\nu} \left( \frac{1}{z}\frac{\mathrm{d}}{\mathrm{d} z} \right)^{\nu} \mathrm{J}_{d,0}(z).
\end{align}

\section{\label{AppendixC} Proof of the property of differential operators induced by spherical harmonics}
Proofs for two and three dimensions can be found in \cite{martin2006,bilbao2019l}. The following argument provides a dimension-independent proof.  
From Eq.~(\ref{eq:relation_RK_JY}), the expansion Eq.~(\ref{eq:SH_expansion}) can be rewritten as
\begin{equation}
    \hat{p}(\bm{r}) = \sum_{\nu=1}^{\infty} \sum_{\mu = -\nu}^{\nu} (-k)^{-\nu} \hat{p}_{\nu}^{\mu} y_{\nu}^{\mu}(\partial_{1})\kappa_{k}(\bm{r}, \bm{0}).
\end{equation}
Using Eq.~(\ref{eq:RK_surface}), one needs to compute
\begin{equation}
    \overline{y_{\nu^{\prime}}^{\mu^{\prime}}}(D) \hat{p}(\bm{0}) =  \sum_{\nu=1}^{\infty} \sum_{\mu = -\nu}^{\nu} (-k)^{-\nu} \hat{p}_{\nu}^{\mu} \left( \overline{y_{\nu^{\prime}}^{\mu^{\prime}}} \cdot y_{\nu}^{\mu} \right) (D) \mathrm{J}_{d,0} (0).
\end{equation}
From Eq.~(\ref{eq:relation_Y_y}), the product $\overline{y_{\nu^{\prime}}^{\mu^{\prime}}} \cdot y_{\nu}^{\mu}$ can be expanded as
\begin{align}
    \overline{y_{\nu^{\prime}}^{\mu^{\prime}}} \cdot y_{\nu}^{\mu} (\bm{r}) &= |\bm{r}|^{\nu^{\prime} + \nu} \overline{Y_{\nu^{\prime}}^{\mu^{\prime}}} \cdot Y_{\nu}^{\mu} \left( \frac{\bm{r}}{|\bm{r}|} \right) \notag \\
    &= |\bm{r}|^{\nu^{\prime} + \nu} \sum_{\nu^{\prime \prime}=0}^{\nu^{\prime} + \nu} \sum_{\mu^{\prime \prime}=-\nu}^{\nu^{\prime \prime}} \langle \overline{Y_{\nu^{\prime}}^{\mu^{\prime}}} \cdot Y_{\nu}^{\mu}, Y_{\nu^{\prime \prime}}^{\mu^{\prime \prime}} \rangle Y_{\nu^{\prime \prime}}^{\mu^{\prime \prime}} \left( \frac{\bm{r}}{|\bm{r}|} \right) \notag \\
    &= \sum_{\nu^{\prime \prime}=0}^{\nu^{\prime} + \nu} \sum_{\mu^{\prime \prime}=-\nu}^{\nu^{\prime \prime}} |\bm{r}|^{\nu^{\prime} + \nu - \nu^{\prime \prime}} \langle \overline{Y_{\nu^{\prime}}^{\mu^{\prime}}} \cdot Y_{\nu}^{\mu}, Y_{\nu^{\prime \prime}}^{\mu^{\prime \prime}} \rangle y_{\nu^{\prime \prime}}^{\mu^{\prime \prime}} (\bm{r}).
\end{align}
Using the integral representation of the reproducing kernel [Eq.~(\ref{eq:kappa_k_integral})] and Eq.~(\ref{eq:J_dn}), one finds
\begin{equation}
    \mathrm{J}_{d,\nu} (0) = \begin{cases}
        |S^{d-1}| & (\nu = 0) \\
        0  & (\nu \neq 0)
    \end{cases},
\end{equation}
so that $y_{\nu^{\prime\prime}}^{\mu^{\prime\prime}} (D) \mathrm{J}_{d,0} (0) = 0$ for $\nu^{\prime\prime} \neq 0$.  
Therefore only the term with $\nu^{\prime\prime}=\mu^{\prime\prime}=0$ survives, and since
\begin{equation}
    \langle \overline{Y_{\nu^{\prime}}^{\mu^{\prime}}} \cdot Y_{\nu}^{\mu}, 1 \rangle = \langle Y_{\nu}^{\mu}, Y_{\nu^{\prime}}^{\mu^{\prime}} \rangle = \delta_{\nu, \nu^{\prime}} \delta_{\mu, \mu^{\prime}},
\end{equation}
and $Y_{0}^{0} = |S^{d-1}|^{-1/2}$, we obtain
\begin{align}
    \overline{y_{\nu^{\prime}}^{\mu^{\prime}}}(D) \hat{p}(\bm{0}) &=  \sum_{\nu=1}^{\infty} \sum_{\mu = -\nu}^{\nu} (-k)^{-\nu} \hat{p}_{\nu}^{\mu} |S^{d-1}|^{-1} \delta_{\nu, \nu^{\prime}} \delta_{\mu, \mu^{\prime}} \Delta^{\nu^{\prime}} \mathrm{J}_{d,0} (0) \notag \\
    &= (-k)^{-\nu^{\prime}} (-k^{2})^{\nu^{\prime}} \hat{p}_{\nu^{\prime}}^{\mu^{\prime}} = k^{\nu^{\prime}} \hat{p}_{\nu^{\prime}}^{\mu^{\prime}}.
\end{align}
This is the desired identity. In the derivation we used that the RK satisfies the Helmholtz equation $(\Delta + k^{2})\mathrm{J}_{d,0} = 0$.


\bibliography{ref}

\begin{thebibliography}{10}
\def\enquote#1,{``#1,''}
\def\enxquote#1{``#1''}
\expandafter\ifx\csname url\endcsname\relax
  \def\url#1{\texttt{#1}}\fi
\expandafter\ifx\csname urlprefix\endcsname\relax\def\urlprefix{URL }\fi
\providecommand{\bibinfo}[2]{#2}
\def\plainquote#1{``#1''}
\providecommand{\noopsort}[1]{}
\providecommand{\switchargs}[2]{#2#1}
\providecommand{\dourl}[1]{\href{http://#1}{\nolinkurl{#1}}}
  \def\eatspace #1{#1}

\bibitem{van2002}
\bibinfo{author}{H.~L. Van~Trees}, \emph{\bibinfo{title}{Optimum array processing: Part IV of detection, estimation, and modulation theory}}  (\bibinfo{publisher}{John Wiley \& Sons}, \bibinfo{address}{Hoboken}, \bibinfo{year}{2002}).

\bibitem{capon1969}
\bibinfo{author}{J.~Capon}, \enquote{\bibinfo{title}{High-resolution frequency-wavenumber spectrum analysis}},  \bibinfo{journal}{Proceedings of the IEEE} \textbf{57}(8), \bibinfo{pages}{1408--1418} (\bibinfo{year}{1969}).

\bibitem{meyer2002}
\bibinfo{author}{J.~Meyer} and \bibinfo{author}{G.~Elko}, \enquote{\bibinfo{title}{A highly scalable spherical microphone array based on an orthonormal decomposition of the soundfield}}, in \emph{\bibinfo{booktitle}{Proc. IEEE Int. Conf. Acoustics, Speech, and Signal Processing (ICASSP)}}, \bibinfo{organization}{IEEE} (\bibinfo{year}{2002}), Vol.~\bibinfo{volume}{2}, pp. \bibinfo{pages}{II--1781--II--1784}.

\bibitem{boaz2015}
\bibinfo{author}{B.~Rafaely}, \emph{\bibinfo{title}{Fundamentals of spherical array processing}}, Vol.~\bibinfo{volume}{8}  (\bibinfo{publisher}{Springer-Verlag}, \bibinfo{address}{Berlin}, \bibinfo{year}{2015}).

\bibitem{ueno2018}
\bibinfo{author}{N.~Ueno}, \bibinfo{author}{S.~Koyama}, and \bibinfo{author}{H.~Saruwatari}, \enquote{\bibinfo{title}{Kernel ridge regression with constraint of helmholtz equation for sound field interpolation}}, in \emph{\bibinfo{booktitle}{Proc. 16th Int. Workshop on Acoustic Signal Enhancement (IWAENC)}} (\bibinfo{year}{2018}), pp. \bibinfo{pages}{436--440}.

\bibitem{iwami2023jasa}
\bibinfo{author}{T.~Iwami}, \bibinfo{author}{K.-i. Sawai}, and \bibinfo{author}{A.~Omoto}, \enquote{\bibinfo{title}{Direction-of-arrival estimation in half-space from single sample array snapshot}},  \bibinfo{journal}{J. Acoust. Soc. Am.} \textbf{153}(5), \bibinfo{pages}{3025--3035} (\bibinfo{year}{2023}).

\bibitem{iwami2024jasa}
\bibinfo{author}{T.~Iwami} and \bibinfo{author}{A.~Omoto}, \enquote{\bibinfo{title}{Frequency-domain sound field from the perspective of band-limited functions}},  \bibinfo{journal}{J. Acoust. Soc. Am.} \textbf{156}(5), \bibinfo{pages}{3298--3305} (\bibinfo{year}{2024}).

\bibitem{bauer1987}
\bibinfo{author}{B.~B. Bauer}, \enquote{\bibinfo{title}{A century of microphones}},  \bibinfo{journal}{J. Audio Eng. Soc.} \textbf{35}(4), \bibinfo{pages}{246--258} (\bibinfo{year}{1987}).

\bibitem{williams1999}
\bibinfo{author}{E.~G. Williams}, \emph{\bibinfo{title}{Fourier acoustics: sound radiation and nearfield acoustical holography}}  (\bibinfo{publisher}{Academic Press}, \bibinfo{address}{San Diego}, \bibinfo{year}{1999}).

\bibitem{balmages2007}
\bibinfo{author}{I.~Balmages} and \bibinfo{author}{B.~Rafaely}, \enquote{\bibinfo{title}{Open-sphere designs for spherical microphone arrays}},  \bibinfo{journal}{IEEE Trans. Audio Speech Lang. Process.} \textbf{15}(2), \bibinfo{pages}{727--732} (\bibinfo{year}{2007}).

\bibitem{poletti2005}
\bibinfo{author}{M.~A. Poletti}, \enquote{\bibinfo{title}{Three-dimensional surround sound systems based on spherical harmonics}},  \bibinfo{journal}{J. Audio Eng. Soc.} \textbf{53}(11), \bibinfo{pages}{1004--1025} (\bibinfo{year}{2005}).

\bibitem{kashiwazaki2022}
\bibinfo{author}{H.~Kashiwazaki}, \bibinfo{author}{T.~Iwami}, and \bibinfo{author}{A.~Omoto}, \enquote{\bibinfo{title}{Axisymmetric directional microphone model for spherical harmonic decomposition of sound field}},  \bibinfo{journal}{Acoust. Sci. \& Tech.} \textbf{43}(3), \bibinfo{pages}{205--208} (\bibinfo{year}{2022}).

\bibitem{bilbao2019i}
\bibinfo{author}{S.~Bilbao}, \bibinfo{author}{J.~Ahrens}, and \bibinfo{author}{B.~Hamilton}, \enquote{\bibinfo{title}{Incorporating source directivity in wave-based virtual acoustics: Time-domain models and fitting to measured data}},  \bibinfo{journal}{J. Acoust. Soc. Am.} \textbf{146}(4), \bibinfo{pages}{2692--2703} (\bibinfo{year}{2019}).

\bibitem{bilbao2019l}
\bibinfo{author}{S.~Bilbao}, \bibinfo{author}{A.~Politis}, and \bibinfo{author}{B.~Hamilton}, \enquote{\bibinfo{title}{Local time-domain spherical harmonic spatial encoding for wave-based acoustic simulation}},  \bibinfo{journal}{IEEE Signal Process. Lett.} \textbf{26}(4), \bibinfo{pages}{617--621} (\bibinfo{year}{2019}).

\bibitem{grochenig2001}
\bibinfo{author}{K.~Gr{\"o}chenig}, \emph{\bibinfo{title}{Foundations of Time-Frequency Analysis}}  (\bibinfo{publisher}{Springer Science \& Business Media, New York}, \bibinfo{year}{2001}).

\bibitem{axler2013}
\bibinfo{author}{S.~Axler}, \bibinfo{author}{P.~Bourdon}, and \bibinfo{author}{R.~Wade}, \emph{\bibinfo{title}{Harmonic function theory}}, Vol. \bibinfo{volume}{137}  (\bibinfo{publisher}{Springer Science \& Business Media}, \bibinfo{address}{New York}, \bibinfo{year}{2013}).

\bibitem{iwami2025ast}
\bibinfo{author}{T.~Iwami}, \bibinfo{author}{N.~Inoue}, and \bibinfo{author}{A.~Omoto}, \enquote{\bibinfo{title}{Orthonormality of spherical basis functions for interior problems of the helmholtz equation}},  \bibinfo{journal}{Acoust. Sci. \& Tech.} \textbf{46}(5), \bibinfo{pages}{590--593} (\bibinfo{year}{2025}).

\bibitem{bowman1958}
\bibinfo{author}{F.~Bowman}, \emph{\bibinfo{title}{Introduction to Bessel functions}}  (\bibinfo{publisher}{Courier Corporation}, \bibinfo{address}{Mineola}, \bibinfo{year}{1958}).

\bibitem{hansen2010}
\bibinfo{author}{P.~C. Hansen}, \emph{\bibinfo{title}{Discrete inverse problems: insight and algorithms}}  (\bibinfo{publisher}{SIAM}, \bibinfo{address}{Philadelphia, PA}, \bibinfo{year}{2010}).

\bibitem{nomura2018}
\bibinfo{author}{T.~Nomura}, \enquote{\bibinfo{title}{A proof of hobson's formula with the euler operator}},  \bibinfo{journal}{Kyushu J. Math.} \textbf{72}(2), \bibinfo{pages}{423--427} (\bibinfo{year}{2018}).

\bibitem{martin2006}
\bibinfo{author}{P.~A. Martin}, \emph{\bibinfo{title}{Multiple scattering: interaction of time-harmonic waves with N obstacles}}, \bibinfo{number}{107}  (\bibinfo{publisher}{Cambridge University Press}, \bibinfo{address}{Cambridge}, \bibinfo{year}{2006}).

\bibitem{mclean2000}
\bibinfo{author}{W.~C.~H. McLean}, \emph{\bibinfo{title}{Strongly elliptic systems and boundary integral equations}}  (\bibinfo{publisher}{Cambridge university press}, \bibinfo{address}{Cambridge}, \bibinfo{year}{2000}).

\bibitem{andrews1998}
\bibinfo{author}{L.~C. Andrews}, \emph{\bibinfo{title}{Special functions of mathematics for engineers}}, Vol.~\bibinfo{volume}{49}  (\bibinfo{publisher}{SPIE Press}, \bibinfo{address}{Bellingham}, \bibinfo{year}{1998}).

\end{thebibliography}


\end{document}